\documentclass[11pt,prd,preprint,superscriptaddress,amsmath,amssymb]{revtex4}
\usepackage{graphicx,color,slashed}

\usepackage{subfigure,bbold}

\def\be{\begin{eqnarray}}
\def\ed{\end{eqnarray}}
\def\non{\nonumber}
\def\ga{\gamma}
\def\lam{\lambda}
\def\beq{\begin{eqnarray}}
\def\eeq{\end{eqnarray}}
\def\beq{\begin{equation}}
\def\eeq{\end{equation}}
\def\be{\begin{eqnarray}}
\def\ed{\end{eqnarray}}
\def\non{\nonumber}
\def\ga{\gamma}

\newcommand{\ba}{\begin{array}}
\newcommand{\ea}{\end{array}}

\begin{document}

\title{\boldmath Two-Higgs-Doublet type-II and -III models and $t\to c h$ at the LHC}

\author{A.~Arhrib \footnote{aarhrib@ictp.it}}
\affiliation{D\'{e}partement de Math\'{e}matiques, 
Facult\'{e} des Sciences et Techniques,
Universit\'{e} Abdelmalek Essaadi, B. 416, Tangier, Morocco.}
\affiliation{Physics Division, National Center for Theoretical Sciences,
  Hsinchu 300, Taiwan}
\author{  R.~Benbrik \footnote{Email: rbenbrik@ictp.it}}
\affiliation{LPHEA, Semlalia, Cadi Ayyad University, Marrakech, Morocco.}
\affiliation{MSISM Team, Facult\'e Polydisciplinaire de Safi, Sidi Bouzid B.P 4162, 46000 Safi, Morocco}
\author{ Chuan-Hung Chen \footnote{physchen@mail.ncku.edu.tw} }
\affiliation{Department of Physics, National Cheng-Kung University, Tainan 70101, Taiwan }
\author{Melina Gomez-Bock \footnote{melina.gomez@udlap.mx}}
\affiliation{DAFM, Universidad de las Am\'ericas
  Puebla. Ex. Hda. Sta. Catarina M\'artir 72810, Cholula, Pue. M\'exico}
\author{Souad Semlali \footnote{s.seemlali@gmail.com}}
\affiliation{LPHEA, Semlalia, Cadi Ayyad University, Marrakech, Morocco}

\date{\today}

\begin{abstract}
  We study the constraints of the generic two-Higgs-doublet model (2HDM)
  type-III and the impacts of the new Yukawa couplings. For comparisons, we
  revisit the analysis in the 2HDM type-II. To understand the influence of  all
  involving free parameters and to realize their correlations, we employ
  $\chi$-square fitting approach by including theoretical and experimental
  constraints, such as S, T, and U oblique parameters, the production of
  standard model Higgs  and its decay to $\gamma\gamma$, $WW^*/ZZ^*$, 
$\tau^+\tau^-$, etc. The errors of analysis are taken at $68\%$, $95.5\%$, and
  $99.7\%$ confidence levels. Due to the new Yukawa couplings being associated with $\cos(\beta-\alpha)$ and $\sin(\beta -\alpha)$, we find that the allowed regions for $\sin\alpha$ and $\tan\beta$ in the type-III model can be broader when the dictated parameter $\chi_F$ is positive; however, for negative $\chi_F$, the limits are  stricter than those in the type-II model. By using the constrained parameters, we find that the deviation from the SM in the $h\to Z\gamma$  can be of ${\cal O}(10\%)$. Additionally, we also study the top-quark flavor-changing processes induced at the tree level in the type-III model and find that  when all current experimental data are considered, we get $Br(t\to c(h, H) )< 10^{-3}$ for $m_h=125.36$ and $m_H=150$ GeV and  $Br(t\to cA)$ slightly exceeds $10^{-3}$ for $m_A =130$ GeV.

\end{abstract}

\maketitle
\section{Introduction}
\label{sec:intro}
A scalar boson around 125 GeV was observed in 2012 by
ATLAS~\cite{:2012gk} and CMS~\cite{:2012gu} at CERN with more than 
$5\sigma$ significance. The discovery of such particle was based on the analyses of following channels: 
$\gamma\gamma$, $WW^*$, $ZZ^*$ and $\tau^+\tau^-$ with  errors of order of  
20-30\% and   $b\bar{b}$ channel with an error of order of 40-50\%.  
The recent updates from ATLAS and CMS with $7\oplus 8$ TeV data~\cite{ATLAS24,ATLAS2G,atlas034,CMS2G,CMS24}
indicate the possible deviations from the standard model (SM) predictions.  Although the errors of current data are still somewhat large,  the new physics signals may become clear in the second run of the LHC at 13-14 TeV.

It is expected that the Higgs couplings to gauge bosons (fermions) at the LHC
indeed could reach  4-6\% (6-13\%) accuracy when the collected data are up to
the integrated luminosity of  300 fb$^{-1}$~\cite{accuracy1,accuracy2}. Furthermore, $e^+ e^-$ Linear Collider (LC) would be able to measure the Higgs couplings at the percent level~\cite{accuracy3}. Therefore, the goals of LHC at run II are (a) to pin down the nature of the observed scalar and see if it is the SM Higgs boson or  a new scalar boson; (b) to reveal the existence of new physics effects, such as the measurement of flavor-changing neutral currents (FCNCs) at the top-quark decays, i.e.  $t\to q h$.

Motivated by the observations of the diphoton, $WW^*$, $ZZ^*$, and
$\tau^+\tau^-$ processes at  the ATLAS and CMS, it is interesting  to investigate
what sorts of models may naturally be consistent with these measurements and
what the implications are  for other channels, e.g. $h\to  Z \gamma$ and $t\to
ch$. Although many possible extensions of the SM have been discussed~\cite{Arhrib:2012yv,Chiang:2012qz}, it is interesting to study the simplest extension from one Higgs doublet to two-Higgs-doublet model (2HDM)
\cite{Lee:1973iz,Branco:2011iw,Ferreira:2011aa,Chiang:2013ixa,Ferreira:2013qua,Barger:2013ofa,Wang:2013sha,Gunion}.  
 According to the situation of Higgs fields coupling to fermions, the 2HDMs are classified 
as type-I, -II, and -III models,  lepton specific model, and flipped model. The 2HDM type-III is the case  where both Higgs doublets couple to all fermions; as a result, FCNCs at the tree level appear. The detailed discussions on the 2HDMs are shown elsewhere~\cite{Branco:2011iw}. 

After scalar particle of 125 GeV is discovered, the implications  of the
observed $h \to \ga\ga$ in the type-I and II models are studied~\cite{Celis:2013rcs} and  the impacts  on  $h \to \ga Z$ are given~\cite{joao,aawh}. As known that the  $\tan\beta$  and angle $\alpha$ are
important free parameters in the 2HDMs, where the former is the ratio of two
vacuum expectation values (VEVs) of Higgses and the latter is the mixing
parameter between the two CP-even scalars.  It is found that the current LHC
data put rather severe constraints on the free parameters~\cite{Ferreira:2011aa}.
For instance, the large $\tan\beta\sim m_t/m_b$ scenario in the type-I and -II is
excluded except if we tune the $\alpha$ parameter to be rather small $\alpha <
0.02$.  Nevertheless,  both type-I and type-II models can still fit the data in some small regions of  $\tan\beta$ and $\alpha$. 

In this paper,  we will explore the influence of  new Higgs couplings
on the $h\to \tau^+\tau^-$, $h\to gg,\ga\ga, WW, ZZ$ and $h\to Z\ga$ decays in the framework of the 2HDM type-III. We will show what is the most favored regions of the type-III parameter space when  theoretical and experimental 
constraints are considered simultaneously. FCNCs of heavy quark such as $t\to
qh$ have been intensively studied both from the experimental and theoretical
point of view~\cite{rev}. Such processes are  well established in the SM and
are excellent probes for the existence of new physics. In the SM and 2HDM type-I and -II, the top-quark  FCNCs are generated at one-loop level by charged currents and are highly suppressed due to the GIM mechanism.  The branching ratio
(BR) for $t\to ch$ in the SM is estimated to be  $3 \times 10^{-14}$
\cite{sm1}. If this decay $t\to ch$  is observed, it would be an indisputable
sign of new physics. Since the tree-level FCNCs appear in the type-III
model, we  explore if  the $Br(t\to ch)$  reaches the order of
$10^{-5}$--$10^{-4}$ \cite{ATLAStch,CMStch}, the sensitivity which is expected
by  the integrated luminosity of 3000 fb$^{-1}$.

The paper is organized as follows. In section II, we introduce the 
scalar potential and the Yukawa interactions in the 2HDM type-III. The theoretical
and experimental constraints are described in section III. We set up the free parameters and establish the $\chi$-square for the best-fit approach in  section VI. In the same section, we discuss the numerical results when all theoretical and experimental constraints are taken into account.  
The conclusions are given in section V.

\section{Model}

In this section we define  the scalar potential and the Yukawa sector in the 2HDM type-III.  The scalar potential in $SU(2)_L\otimes U(1)_Y$ gauge symmetry and CP invariance is given by \cite{Gunion:2002zf}
\be
V(\Phi_1,\Phi_2)
&=& m^2_1 \Phi^{\dagger}_1\Phi_1+m^2_2 \Phi^{\dagger}_2\Phi_2 -(m^2_{12}
\Phi^{\dagger}_1\Phi_2+{\rm h.c}) +\frac{1}{2} \lam_1 (\Phi^{\dagger}_1\Phi_1)^2 
\nonumber \\ &+& \frac{1}{2} \lam_2
(\Phi^{\dagger}_2\Phi_2)^2 +
\lam_3 (\Phi^{\dagger}_1\Phi_1)(\Phi^{\dagger}_2\Phi_2) + \lam_4
(\Phi^{\dagger}_1\Phi_2)(\Phi^{\dagger}_1\Phi_2)  \non \\
&+& 
 \left[\frac{\lam_5}{2}(\Phi^{\dagger}_1\Phi_2)^2 +  \left(\lam_6 \Phi^\dagger_1 \Phi_1 + \lam_7 \Phi^\dagger_2 \Phi_2 \right) \Phi^\dagger_1 \Phi_2+{\rm h.c.} \right] ~,
\label{higgspot}
\ed
where the doublets $\Phi_{1,2}$ have weak hypercharge $Y=1$, the corresponding VEVs are $v_1$ and $v_2$, and
$\lambda_i$ and $m^2_{12}$ are real parameters.  After electroweak symmetry breaking, three of the eight degrees
of freedom in the two Higgs doublets are the Goldstone
bosons ($G^\pm$, $G^0$) and the remaining five degrees of freedom become the
physical Higgs bosons: 2 CP-even $h$, $H$, one CP-odd $A$, and a pair of
charged Higgs  $H^\pm$. After using the minimized conditions and the W mass,
the potential in Eq.~(\ref{higgspot}) has  nine parameters  which will be taken as:
$(\lambda_i)_{i=1,\ldots,7}$, $m^2_{12}$, and $\tan\beta \equiv
v_2/v_1$. Equivalently, we can use the masses as the independent parameters; therefore, the set of free parameters can be chosen to be
\begin{equation}
\{ m_{h}\,, m_{H}\,, m_{A}\,, m_{H^\pm}\,, \tan\beta\,,  \alpha\,,  m^2_{12} \}\,, 
\label{parameters} 
\end{equation}
where we only list seven of the nine parameters, the angle $\beta$ diagonalizes
the squared mass matrices of CP-odd and charged scalars and the angle $\alpha$
diagonalizes the CP-even squared-mass matrix. In order to avoid generating spontaneous CP violation, we further require
 \be
 m^2_{12} - \frac{\lambda_6 v^2_1}{2} - \frac{\lambda_7 v^2_2}{2} \geq \zeta \lambda_5 v_1 v_2
 \ed
 with $\zeta=1(0)$ for $\lam_5> (< )0$~\cite{Gunion:2002zf}.
It has been known that by assuming neutral flavor conservation at the tree-level \cite{Glashow:1976nt},  we have four types of Higgs couplings to the fermions. In the 2HDM type-I, the quarks and leptons couple only to one of the two Higgs doublets and the case  is the same as the SM.
In the 2HDM type-II, the charged leptons and down type quarks couple to one Higgs doublet and the up type quarks couple to the other. 
The lepton-specific model is similar to type-I, but the leptons couple to the other Higgs doublet. In the flipped model, 
which is similar to type-II, the leptons and up type quarks couple to the same double.

If the tree-level FCNCs are allowed, both doublets can couple to leptons and
quarks  and the associated model is called 2HDM type-III~\cite{Cheng:1987rs,Atwood:1996vj,Branco:2011iw}. Thus, 
the Yukawa interactions for quarks are written as
\beq
{\cal {L}}_{Y} =  \bar Q_{L} 
Y^{k}  d_{R} \phi_k+  \bar Q_{L }  \tilde{Y}^{k}  u_{R} \tilde{\phi}_k + h.c.
\eeq
where the flavor indices are suppressed, $Q^T_L=(u_L, d_L)$ is the left-handed quark 
doublet, $Y^k$ and  $\tilde{Y}^k$ denote the $3\times 3$ 
Yukawa matrices,  $\tilde{\phi_k}=i\sigma_2 \phi^*_k$, and $k$ is the 
doublet number. Similar formulas could be applied to the lepton sector. 
Since the mass matrices of quarks are combined by $Y^{1} (\tilde Y^{1})$ 
and $Y^2 (\tilde Y^{2})$ for down (up) type quarks and  $Y^{1,2}(\tilde Y^{1,2})$ 
generally cannot be diagonalized simultaneously, 
as a result, the tree-level FCNCs appear and the effects lead to the 
oscillations of $K-\bar K$, $B_q-\bar B_q$ and $D-\bar D$ at the tree-level.  
To get naturally small FCNCs, one can use the ansatz formulated by $Y^{k}_{ij},\,\tilde Y^k_{ij} \propto \sqrt{m_i m_j}/v$ 
\cite{Cheng:1987rs,Atwood:1996vj}. After  spontaneous symmetry breaking,  
the scalar couplings to fermions can be expressed as \cite{GomezBock:2005hc}
\be
 {\cal L}^{\rm 2HDM-III}_Y  &=&
\bar u_{Li} \left( \frac{\cos\alpha}{\sin\beta}\frac{m_{u_i}}{v} \delta_{ij} - \frac{\cos(\beta-\alpha)}{\sqrt{2}\sin\beta} X^u_{ij} \right) u_{Rj} h \non \\
&+& \bar d_{Li} \left( -\frac{\sin\alpha}{\cos\beta}\frac{m_{d_i}}{v} \delta_{ij}  + \frac{\cos(\beta-\alpha)}{\sqrt{2}\cos\beta} X^d_{ij} \right) d_{Rj} h \non \\ 
&+& \bar u_{Li} \left( \frac{\sin\alpha}{\sin\beta}\frac{m_{u_i}}{v} \delta_{ij} + \frac{\sin(\beta-\alpha)}{\sqrt{2}\sin\beta} X^u_{ij} \right) u_{Rj} H \non\\
&+& \bar d_{Li} \left(\frac{\cos\alpha}{\cos\beta}\frac{m_{d_i}}{v} \delta_{ij}
 - \frac{\sin(\beta-\alpha)}{\sqrt{2}\cos\beta} X^d_{ij} \right) d_{Rj} H \non\\
&-& i\bar u_{Li} \left(\frac{1}{\tan\beta}\frac{m_{u_i}}{v} \delta_{ij} -
\frac{X^u_{ij} }{\sqrt{2}\sin\beta}  \right) u_{Rj} A \non \\
&+& i\bar d_{Li} \left(-\tan\beta\frac{m_{d_i}}{v} \delta_{ij}
 + \frac{X^d_{ij}}{\sqrt{2}\cos\beta} \right) d_{Rj} A +{\rm h.c}\,,
\label{eq:LhY}
\ed
where $v=\sqrt{v^2_1 + v^2_2}$, 
$X^q_{ij}=\sqrt{m_{q_i} m_{q_j}}/v \chi^q_{ij}$ ($q=u,d$ ) and
$\chi^{q}_{ij}$ are the free parameters. By above formulation, if the FCNC effects are ignored, the results are returned to
the case of the 2HDM type-II, given by
\be
{\cal L}^{2HDM-II}_Y &=& 
\bar u_{Li} \left( \frac{\cos\alpha}{\sin\beta}\frac{m_{u_i}}{v} \delta_{ij}
\right) u_{Rj} h +\bar d_{Li} \left(-\frac{\sin\alpha}{\cos\beta}\frac{m_{d_i}}{v}
\delta_{ij} \right) d_{Rj} h  + {\rm h.c}\,.
\label{eq:LhYII}
\ed

For the couplings of other scalars to fermions,
the can be found elsewhere~\cite{GomezBock:2005hc}. It can be seen clearly that if 
$\chi^{u,d}_{ij}$ are of ${\cal{O}}(10^{-1})$, the new effects are dominated 
by heavy fermions and comparable with those in the type-II model. 
The couplings of $h$ and $H$ to gauge bosons $V=W,Z$ are proportional to 
$\sin(\beta-\alpha)$ and  $\cos(\beta-\alpha)$, respectively.
Therefore, the SM-like Higgs boson $h$ is recovered 
when $\cos(\beta-\alpha)\approx 0$.  The decoupling limit 
can be achieved if $\cos(\beta-\alpha)\approx 0$ and  $m_h \ll m_H, m_A, m_{H\pm}$
are satisfied~\cite{Gunion:2002zf}. From Eqs. (\ref{eq:LhY}) and (\ref{eq:LhYII}), one can also find that in the decoupling limit, the h couplings to quarks are returned to the SM case.

In this analysis, since we take $\alpha$ in the range $-\pi/2 \leq \alpha\leq \pi/2$, 
$\sin\alpha$ will have both positive and negative sign. In the 2HDM type-II, if $\sin\alpha <0$ then the Higgs 
couplings to up- and down-type quarks will have the same sign as those  in the SM. 
It is worthy to mention that $\sin\alpha$ in minimal supersymmetric SM (MSSM) is negative unless some extremely large radiative corrections 
flip its sign \cite{Gunion:2002zf}. If $\sin\alpha$ is positive, then 
the Higgs coupling to down quarks
will have a different sign with respect to the SM case. This is called 
by the  wrong sign Yukawa coupling in the literature~\cite{wrong2,Gunion:2002zf}. 
Later we will explain if the type-III model would favor such 
wrong sign scenario or not.
\section{Theoretical and experimental constraints}

The free parameters in the scalar potential defined in Eq.~(\ref{higgspot}) could be constrained by theoretical requirements
 and the experimental measurements, where the former mainly includes tree level unitarity and vacuum stability
 when the electroweak symmetry is broken spontaneously.  Since the unitarity constraint involves a variety of scattering processes, 
 we adopt the results~\cite{abdesunit,abdesunit1}.
We also force the potential to be 
perturbative by requiring that all quartic couplings of the scalar 
potential obey $|\lambda_i| \leq 8 \pi$ for all $i$. 
For the vacuum stability conditions which ensure that the potential 
is bounded from below, we require that the parameters satisfy the conditions as ~\cite{vac1,vac2}
\begin{eqnarray}
\nonumber
&& \lambda_1  > 0\;,\quad\quad \lambda_2 > 0\;, \lambda_3 +
  \sqrt{\lambda_1\lambda_2 } > 0, \quad 
 \sqrt{\lambda_1\lambda_2 }
+ \lambda_{3} + \lambda_{4}  -|\lambda_{5}| >0,\\ &&
{2|\lambda_6 + \lambda_7| \le \frac{1}{2}(\lambda_1 + \lambda_2) +
  \lambda_3 + \lambda_4 + \lambda_5}\,.
\label{vac}
\end{eqnarray}

In the following we state the constraints from experimental data. The new neutral and charged scalar bosons in 2HDM will affect the self-energy of W and Z bosons through the loop effects. Therefore, the involved parameters could be constrained by the precision measurements of the oblique parameters, denoted by S, T, and U~\cite{Peskin}. Taking $m_h = 125$ GeV,  $m_t = 173.3$ GeV and assuming that $U=0$,  the tolerated ranges for S and T are found by~\cite{Baak:2014ora}
\begin{eqnarray}
 \Delta S = 0.06\pm0.09\,, \ \  \Delta T = 0.10\pm0.07\,,
\label{test:ST}
\end{eqnarray}
where the correlation factor is $\rho=+0.91$, $\Delta S = S^{\textrm{2HDM}} - S^{\textrm{SM}}$ and 
$\Delta T = T^{\textrm{2HDM}} - T^{\textrm{SM}}$, and their explicit expressions  can be found~{\cite{Gunion:2002zf}}. We note that in the limit $m_{H^\pm}=m_{A^0}$ or $m_{H^\pm}=m_{H^0}$,
$\Delta T$ vanishes~\cite{Gerard:2007kn,Cervero:2012cx}. 

The second set of constraints comes from B physics observables.
It has been shown recently  in Ref.~\cite{Misiak:2015xwa} that 
 $Br(\overline{B}\to  X_s \gamma)$ gives a lower limit 
on $m_{H^\pm}\geq 480$ GeV in the type-II  at $95\%$CL. 
 By the precision measurements of $Z\to b \bar{b}$ and $B_q - \bar{B}_q$ mixing, the values of $\tan\beta < 0.5$ have been excluded~\cite{Baak:2011ze}. In this work we allow $\tan\beta \geq 0.5$. Except some  specific scenarios, $\tan \beta$ can not be too large due
 to the requirement of perturbation theory.

 By the observation of scalar boson at $m_h \approx 125$ GeV,  the searches for Higgs boson at ATLAS and CMS can give strong bounds on the free parameters.  The signal events in the Higgs measurements are represented by the signal strength, which is defined by the ratio of Higgs signal to the SM prediction and  given by
\begin{eqnarray}
\mu^f_i= \frac{\sigma_i(h) \cdot Br(h\to f) }{\sigma^{SM}_i(h)
\cdot Br^{SM}(h\to f) } \equiv \bar \sigma_i \cdot \mu_f \,, \label{eq:kvf}
\end{eqnarray}
where $\sigma_i(h)$ denotes the Higgs production cross section by channel $i$
and $Br(h\to f)$ is the BR for the Higgs decay $h\to
f$. Since several Higgs boson production channels are available at the LHC, we
are interested in the gluon fusion production ($ggF$), $t \bar t h$, vector
boson fusion (VBF) and Higgs-strahlung $Vh$ with $V=W/Z$; and they are grouped
to be $\mu^f_{ggF+t\bar t h}$ and $\mu^f_{VBF+Vh}$.  
In order to consider the constraints from the current LHC data, the  scaling factors which show the Higgs coupling deviations
 from the SM are defined as
\begin{align}
&\kappa_V^{} =\kappa_W = \kappa_Z \equiv 
\frac{g_{hVV}^{\text{2HDM}}}{g_{hVV}^{\text{SM}}},~\quad
\kappa_f^{} \equiv \frac{y_{hff}^{\text{2HDM}}}{y_{hff}^{\text{SM}}},\quad
\label{scaling1}
\end{align} 
where $g_{hVV}$ and $y_{hff}$ are the Higgs couplings to
gauge  bosons and fermions, respectively, and  $f$ stands for top, bottom
quarks, and tau lepton. 
The scaling factors for loop-induced channels are defined by 
\begin{eqnarray}
 \kappa_\gamma^2 &\equiv& \frac{\Gamma(h\to
  \gamma\gamma)_{\text{2HDM}}}{\Gamma(h\to \gamma\gamma)_{\text{SM}}}\,,~\quad \kappa_g^2 \equiv \frac{\Gamma(h\to
  g\,g)_{\text{2HDM}}}{\Gamma(h\to g\,g)_{\text{SM}}}\,, \non \\
   \quad \kappa_{Z\gamma}^2 &\equiv& \frac{\Gamma(h\to
  Z\gamma)_{\text{2HDM}}}{\Gamma(h\to Z\gamma)_{\text{SM}}}\,,~\quad \kappa_{h}^2 \equiv \frac{\Gamma(h)_{\text{2HDM}}}{\Gamma(h)_{\text{SM}}}\,,
\end{eqnarray}
where $\Gamma(h\to XY)$ is the partial decay rate for $h\to XY$.  In this study, the partial decay width of the Higgs is taken from
 \cite{anatomy}, where QCD corrections have been taken into account.
In the decay modes $h\to \ga\ga$ and $h\to Z\ga$,  we have included the contributions  of charged Higgs  and   new Yukawa couplings. Accordingly, the  ratio of cross section to the SM prediction for the production channels $ggF+t\bar t h$ and VBF+$Vh$ can be expressed as
\begin{eqnarray}
\overline{\sigma}_{ggF+t\bar t h} &=& \frac{\kappa^2_{g}\sigma_{SM}(ggF) +
  \kappa^2_{t}\sigma_{SM}(tth)}{\sigma_{SM}(ggF) + \sigma_{SM}(tth)}\,, \quad \\
\overline{\sigma}_{VBF+Vh} &=&
\frac{\kappa^2_{V}\sigma_{SM}(VBF)+\widetilde{\kappa}_{Zh}\widetilde{\sigma}_{SM}(Zh)
  + \kappa^2_{V}\sigma_{SM}(Zh) + \kappa^2_{V}\sigma_{SM}(Wh)}{\sigma_{SM}(VBF)+\widetilde{\sigma}_{SM}(Zh)
  + \sigma_{SM}(Zh) + \sigma_{SM}(Wh)}\,,
\label{Eq:XS1}
\end{eqnarray}
where $\sigma_{SM}(Zh)$ is from the coupling of $ZZh$ and occurs at the tree level and  $\widetilde{\sigma}_{SM}(Zh)\equiv \sigma_{SM} (gg\to Zh)$ represents the  effects of top-quark  loop. With $m_h=125.36$ GeV, the scalar factor $\widetilde{\kappa}_{Zh}$ can be written as~\cite{ATLAS24}
\begin{eqnarray}
\widetilde{\kappa}_{Zh} &=& 2.27\kappa^2_Z + 0.37\kappa^2_t - 1.64\kappa_Z\kappa_t\,.
\end{eqnarray}
In the numerical estimations, we use  $m_h = 125.36$ GeV which is from  LHC Higgs Cross Section Working
Group~\cite{accuracy1} at $\sqrt{s} = 8$ TeV.  The experimental values of signal strengths are shown in Table.~\ref{data:1}, where  the results of ATLAS~\cite{atlas034}
and CMS~\cite{cms005} are combined and denoted by $\widehat{\mu}^f_{ggF+t\bar t h}$ and
$\widehat{\mu}^f_{VBF+Vh}$.
\begin{table}[hptb]  
\begin{center}
\caption{ Measured signal strengths $\widehat{\mu}_{\rm{ggF+tth}}$ and 
$\widehat{\mu}_{\rm{VBF+Vh}}$ that combine the best fit of ATLAS and CMS 
and correlation coefficient $\rho$ for the Higgs decay 
 mode~\cite{atlas034,cms005}.}
\renewcommand{\arraystretch}{1.1}
\begin{tabular}{c||ccccc}
\hline\hline
$f$ & $\widehat{\mu}^{f}_{\rm{ggF+tth}}$ & $\widehat{\mu}^{f}_{\rm{VBF+Vh}}$ &$\pm\,\,1\widehat{\sigma}_{\rm{ggF+tth}}$
&  $\pm\,\,1\widehat{\sigma}_{\rm{VBF+Vh}}$& $\rho$ \\ \hline 
$\gamma\gamma$ & $1.32 $ & 0.8 & 0.38 & 0.7 & -0.30\\ \hline 
$ZZ^*$ & $1.70 $ & 0.3 & 0.4 & 1.20 & -0.59\\ \hline 
$WW^*$ & $0.98 $ & 1.28 & 0.28 & 0.55 & -0.20\\ \hline
$\tau\tau$ & $2 $ & 1.24 & 1.50 & 0.59 & -0.42\\ \hline
$b\bar{b}$ & 1.11 & 0.92 & 0.65 & 0.38 & 0 \\ \hline\hline
\end{tabular}
\label{data:1}
\end{center}
\end{table}
\section{Parameter setting, global fitting, and numerical results}
\subsection{Parameters and global fitting}
After introducing the scaling factors  for displaying  the new physics in various channels, in the following we show the explicit relations with the free parameters in the type-III model.  By the definitions in Eq.~(\ref{scaling1}), the scaling factors for $\kappa_V$ and $\kappa_f$ in the type-III are given by
\begin{eqnarray}
\kappa_V^{} &=& \sin(\beta-\alpha)\,, \non \\ 
\kappa_U^{} &=& \kappa_t = \kappa_c =  \frac{\cos\alpha}{\sin\beta} -
\chi_F \frac{\cos(\beta-\alpha)}{\sqrt{2}\sin\beta}\,,  \non \\
\kappa_D^{} &=& \kappa_b = \kappa_{\tau} = -\frac{\sin\alpha}{\cos\beta} +
\chi_F \frac{\cos(\beta-\alpha)}{\sqrt{2}\cos\beta}\,.
\label{scaling3}
\end{eqnarray}
Although FCNC processes give strict constraints on flavor changing couplings $\chi^{f}_{ij}$ with $i\neq j$, however, the constraints are  applied to the flavor changing processes in $K$, $D$ and $B$ meson systems. Since the couplings of scalars to the light quarks have been suppressed by $m_{q_i}/v$, the direct limit on flavor-conserved coupling $\chi^f_{ii}$ is mild. Additionally, since the signals for top-quark flavor changing processes haven't been observed yet, the direct constraint on $X^u_{3i}= \sqrt{m_t m_{q_i}}/v \chi^u_{3i}$ is from the experimental bound of $t\to h q_i$. Hence, for simplifying the numerical analysis, in Eq.~(\ref{scaling3}) we have set  $\chi^u_{22}=\chi^u_{33} = \chi^{d}_{33} =\chi^{\ell}_{33}=\chi_F$. Since $X^u_{33} = m_t/v \chi_F$, it is conservative to adopt the vale of $\chi_F$ to be ${\cal O}(1)$. In the 2HDM, the charged Higgs will also contribute to  $h\to \gamma \gamma$  decay and the associated  scalar triplet coupling $hH^+ H^-$  is read by 
\begin{eqnarray}
\lambda_{hH^{\pm}H^{\mp}} &=&
\frac{1}{2m^2_W}\Bigg(\frac{\cos(\alpha+\beta)}{\sin2\beta}(2m^2_h - 2\lambda_5
v^2) - \sin(\beta-\alpha)(m^2_h - 2m^2_{H^\pm})\nonumber \\  &+&
m^2_W\cos(\beta-\alpha)\left( \frac{\lambda_6}{\sin^2\beta} - \frac{\lambda_7}{\cos^2\beta}\right)\Bigg)\,. \label{eq:hHH}
\end{eqnarray}
The scaling factors for loop induced processes $h\to (\gamma \gamma, Z\gamma, gg)$ can be expressed by
\begin{eqnarray}
\kappa_{\gamma}^2 & \sim & \Big|1.268 \kappa_W - 0.279\kappa_t + 0.0042\kappa_b + 0.0034\kappa_c +
0.0036\kappa_\tau - 0.0014\lambda_{hH^{\pm}H^{\mp}} \Big|^2\,, \non  \\
\kappa_{Z\gamma}^2 & \sim &\Big|1.058 \kappa_W - 0.059 \kappa_t + 0.00056
\kappa_b + 0.00014 \kappa_c -0.00054 \lambda_{hH^{\pm}H^{\mp}} \Big|^2\,, \non  \\
\kappa_{g}^2  &\sim &  
\Big|1.078 \kappa_t - 0.065\kappa_b - 0.013\kappa_c \Big|^2\,,
\label{scaling2} 
\end{eqnarray}
where we have used $m_h = 125.36$ GeV and  taken $m_{H^\pm} = 480$ GeV. It is
clear that  the charged Higgs contribution to $h\to \gamma\gamma$ and $h\to Z\gamma$ is small. 
In order to study the influence of new free parameters and to understand their correlations, we perform the $\chi$-square fitting  by using  the LHC data for Higgs searches~\cite{:2012gk,:2012gu,ATLAS2G,CMS2G}.
For a given channel $f=\gamma\gamma, W W^*, Z Z^*, \tau \tau$, we define the $\chi^2_f$ as
\begin{equation}
\chi^2_f = \frac{1}{\hat{\sigma}^2_{1}(1-\rho^2)}(\mu^{f}_{1} - \hat{\mu}^{f}_{1})^2
+ \frac{1}{\hat{\sigma}^2_{1}(1-\rho^2)}(\mu^{f}_{2} - \hat{\mu}^{f}_{2})^2 - \frac{2\rho}{\hat{\sigma}_{1}\hat{\sigma}_{2}(1-\rho^2)}(\mu^{f}_{1} - \hat{\mu}^{f}_{1})(\mu^{f}_{2} - \hat{\mu}^{f}_{2})\,, \label{eq:chi2}
\end{equation}
where $\hat{\mu}^{f}_{1,2}$, $\hat{\sigma}_{1,2}$ and $\rho$ are the measured
Higgs signal strengths, their one-sigma errors, and their correlation,
respectively and their values could refer to 
Table~\ref{data:1}, the indices $1$ and $2$ in turn stand for $\rm ggF+tth$ and $\rm VBF+Vh$, and 
$\mu^{f}_{1,2}$  are the results in the 2HDM. The global $\chi$-square  is defined by
\begin{equation}
\chi^2 = \sum_{f}\chi^2_{f} + \chi^2_{ST} \,,
\end{equation}
where the $\chi^2_{ST}$ is related to the $\chi^2$ for S and T parameters, the definition
can be obtained from 
Eq.(\ref{eq:chi2}) by replacing $\mu^f_1 \to S^{2HDM}$ and $\mu^f_2\to T^{2HDM}$, and the corresponding
values can be found from Eq.~(\ref{test:ST}). We do not include $b\bar{b}$ channel in our analysis because the errors of data are still large. 

In order to display the allowed regions for the parameters, we show the best fit at $68\%$, $95.5\%$, and $99.7\%$ confidence levels (CLs), that is, the corresponding errors of $\chi^2$ are $\Delta \chi^2 \leq 2.3$, $5.99$, and $11.8$, respectively. For comparing with LHC data, we require the calculated results  in agreement with those shown in ATLAS Fig.~3 of Ref.~\cite{ATLAS24} and in CMS Fig.~5 of Ref.~\cite{CMS24}.  

\section{Numerical Results}
In the following we present the limits of current LHC data based on the three
kinds of CL introduced in last section. In our numerical calculations, we set
the mass of SM Higgs to be $m_h=125.36$ GeV,  and  scan the involved parameters in the  chosen regions  as 
\begin{eqnarray}
&& 480 \,{\rm GeV} \le m_{H^\pm} \le 1 \,{\rm TeV}, \quad 126 \,{\rm GeV} 
\le m_{H} \le 1 \,{\rm TeV}, 
\quad 100 \,{\rm GeV} \le m_{A} \le 1 \,{\rm TeV} \,, \non \\
&& -1 \le \sin\alpha \le 1,  \quad 0.5 \le \tan\beta \le 50,  
\quad -(1000\, {\rm GeV})^2 \le m^2_{12} \le (1000\, {\rm GeV})^2 \,.
\label{numbers}
\end{eqnarray}

\begin{figure}[h!]
\includegraphics[width=0.32\textwidth]{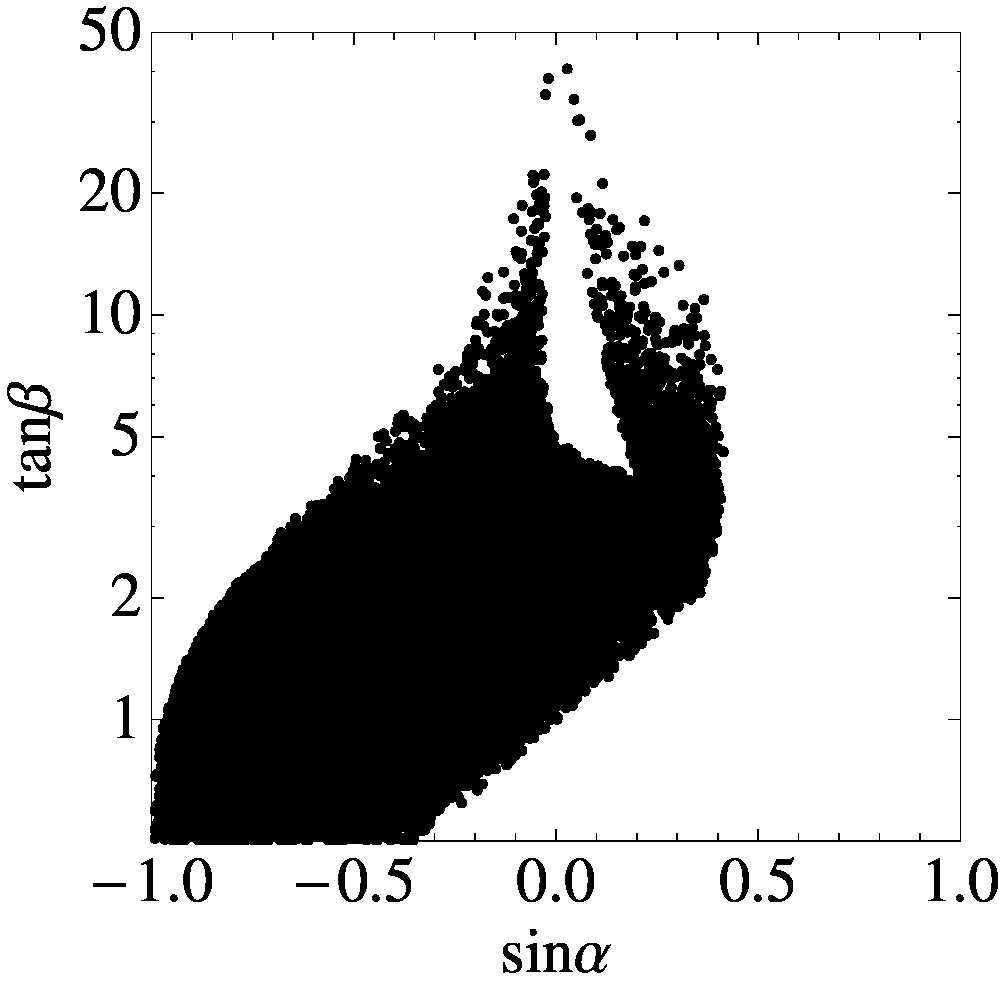}
\includegraphics[width=0.32\textwidth]{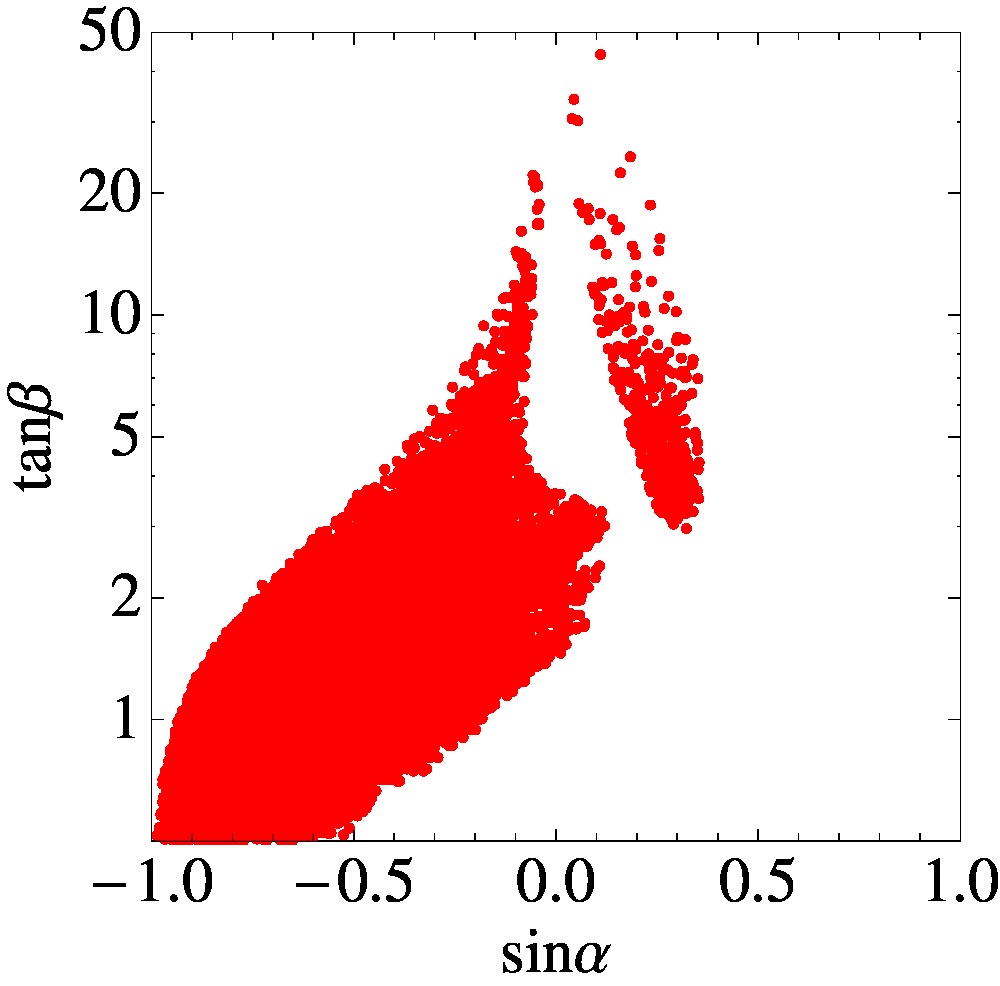}
\includegraphics[width=0.32\textwidth]{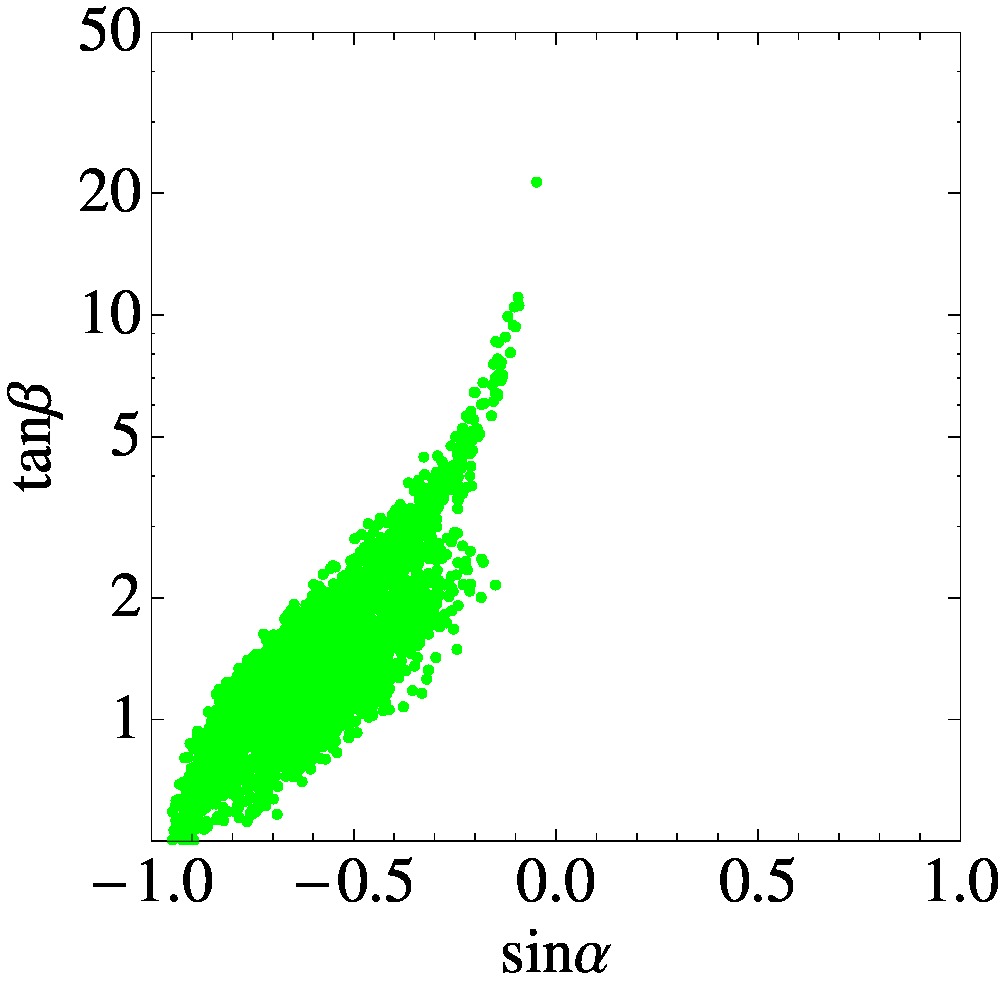}
\caption{The allowed regions in $(\sin\alpha, \tan\beta)$ 
constrained by theoretical and current experimental inputs, where we have used 
$m_h = 125.36 $ GeV in the type-III with $-1\le\chi_F\le 1$ and $-1\le\lambda_{6,7}\le 1$. The
errors for $\chi$-square fit are 99.7$\%$ CL (left panel),  95.5$\%$ CL
(middel panel) and 68$\%$ CL(right panel). }
\label{fig:satb_l67}
\end{figure}

\begin{figure}[hptb]
\includegraphics[width=0.32\textwidth]{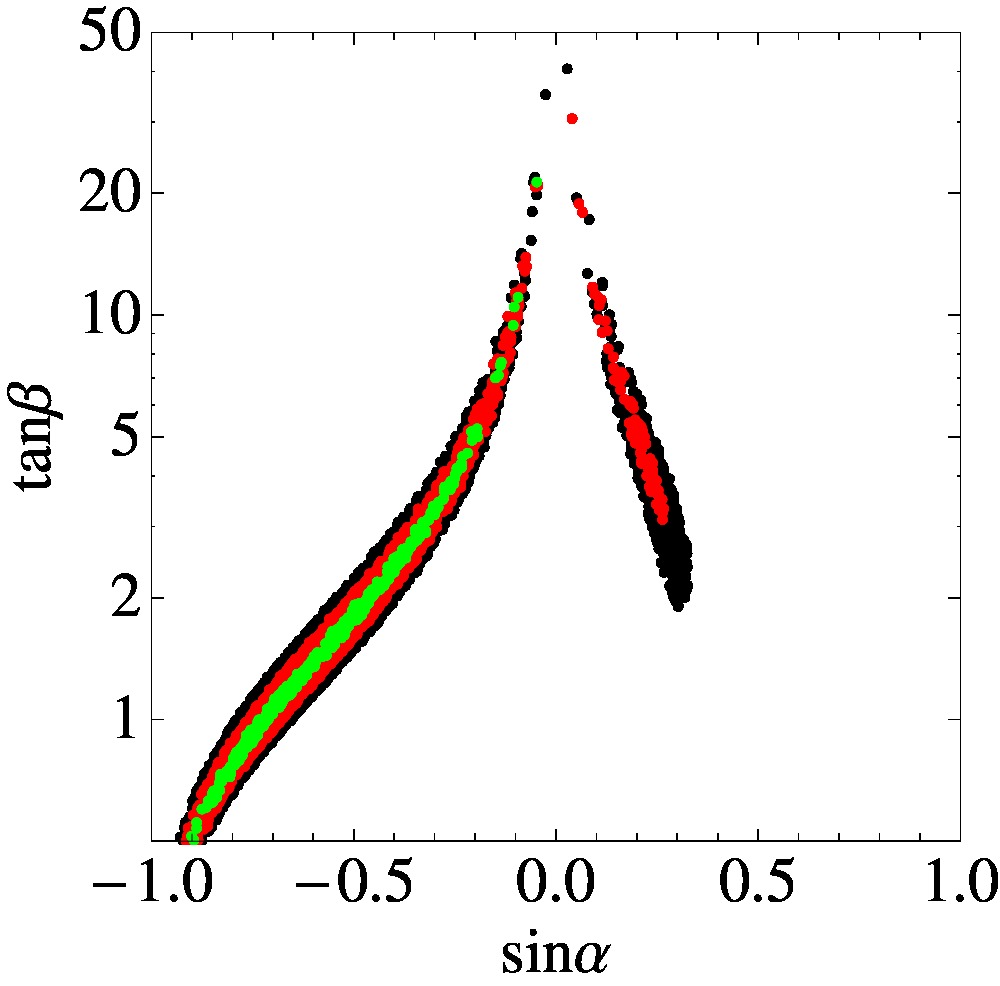}
\includegraphics[width=0.32\textwidth]{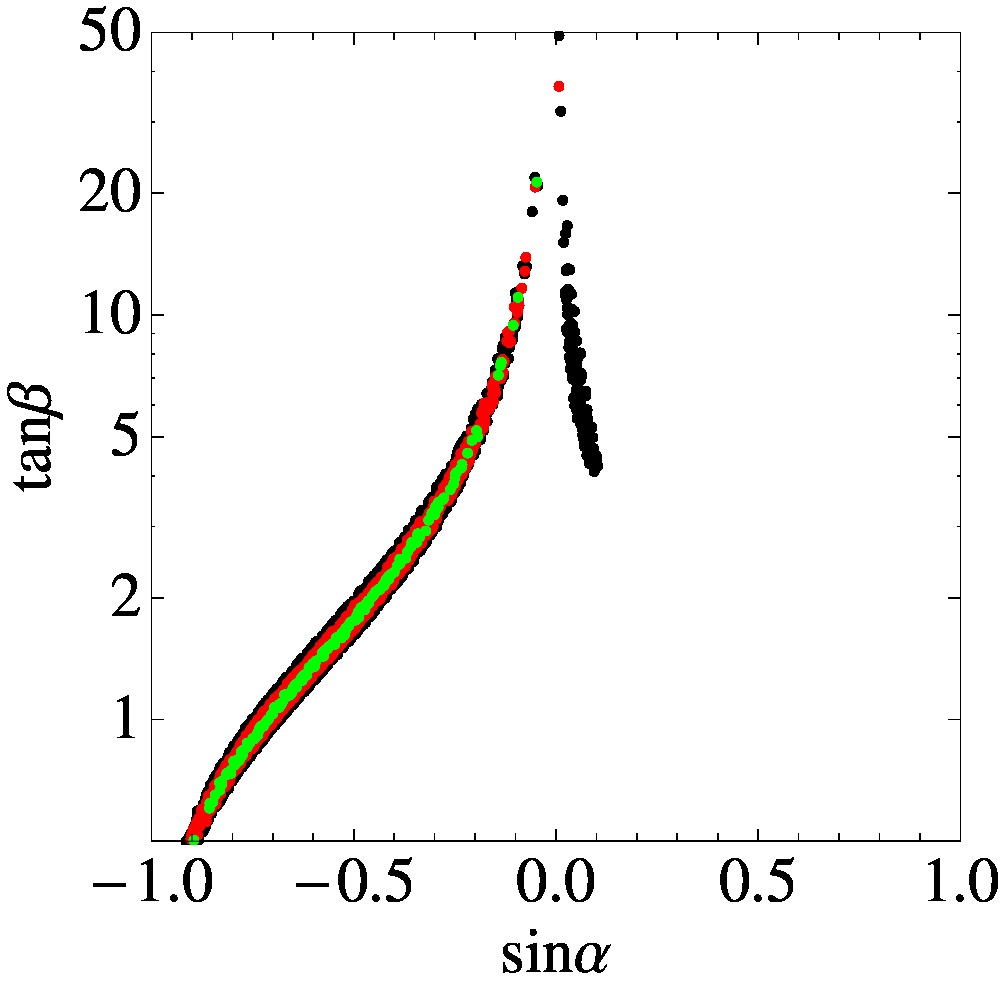}
\includegraphics[width=0.32\textwidth]{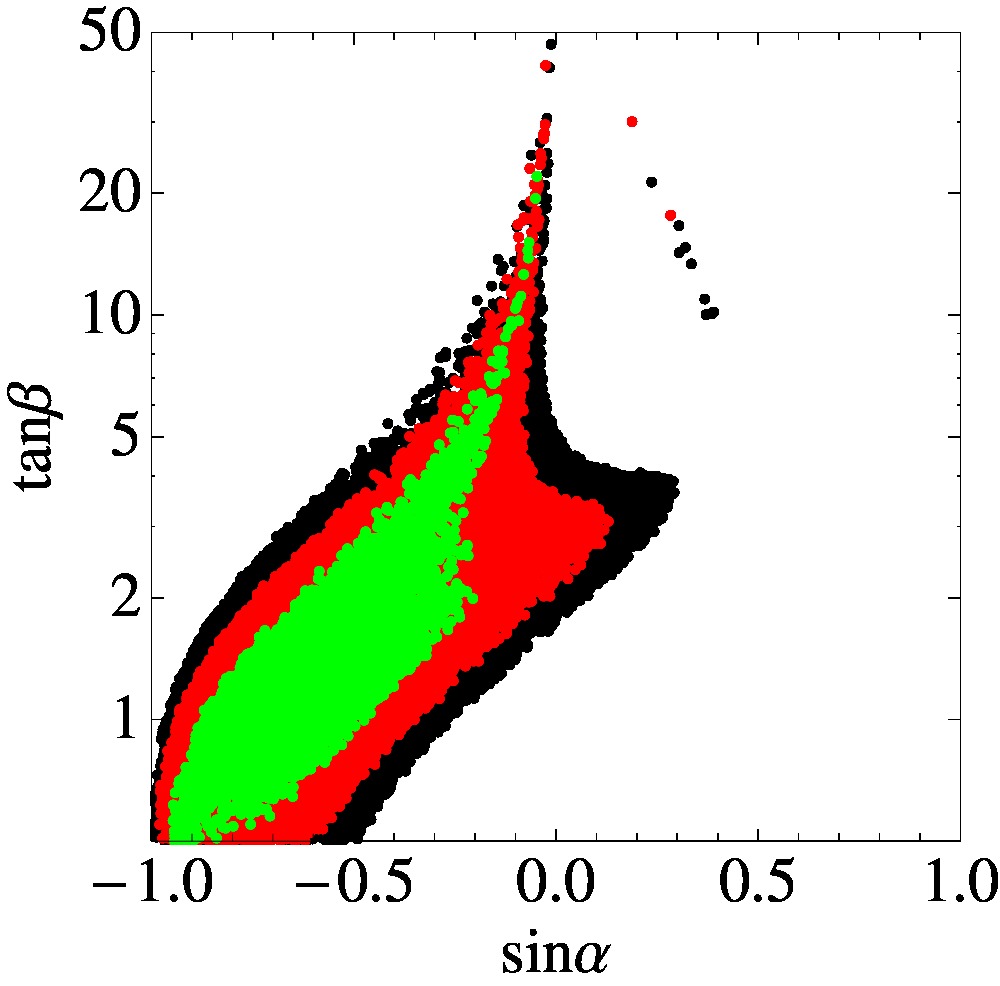}
\caption{The allowed regions in $(\sin\alpha, \tan\beta)$ 
constrained by theoretical and current experimental inputs, where we have used 
$m_h = 125.36 $ GeV, the left, middle and right panels stand for the 2HDM type-II, type-III with $\chi_F=-1$ and type-III with $\chi_F=+1$, respectively.  The errors for $\chi$-square fit are 99.7$\%$ CL (black),  95.5$\%$ CL (red) and 68$\%$ CL  (green). }
\label{fig:satb}
\end{figure}

 The main difference in the scalar potential between 
type-II and type-III is that the  $\lambda_{6,7}$ 
terms appear in the type-III model. With the introduction of 
$\lambda_{6,7}$ terms in the potential, not only the  mass relations of 
scalar bosons are modified but also the scalar triple and quartic coupling 
receives contributions from $\lambda_{6}$ and $\lambda_{7}$.
Since the masses of scalar bosons 
are regarded as free parameters,  the relevant  $\lambda_{6,7}$ effects 
in this study enters game through  the triple coupling 
$h$-$H^+$-$H^-$ that contributes to the 
$h\to \gamma\gamma$ decay, as shown in Eq.~(\ref{eq:hHH}) and the first line of Eq.~(\ref{scaling2}). 
Since the contribution of the charged Higgs loop  to the 
$h\to \gamma\gamma$ decay is small, expectably the influence of 
$\lambda_{6,7}$ on the parameter constraint is not significant. 
To demonstrate that the contributions of $\lambda_{6,7}$ are not 
very important, we present the allowed ranges for $\tan\beta$ and 
$\sin\alpha$ by scanning $\lambda_{6,7}$ in the region of $[-1,1]$ 
in Fig.~\ref{fig:satb_l67}, where the theoretical and experimental 
constraints mentioned earlier are included and the plots from left to 
right in turn stand for $\Delta \chi^2 =11.8$, $5.99$, and $2.3$, 
respectively. Additionally, to understand the influence of $\chi_F$ 
defined in Eq.~(\ref{scaling3}), we also scan $\chi_F=[-1,1]$ in the plots. 
By comparing  the results with the case of $\lambda_{6,7}\ll 1$ 
and $\chi_F= 1$ which is displayed in the third plot of 
Fig.~(\ref{fig:satb}), it can be seen that only small region in 
the positive  $\sin\alpha$ is modified and the modifications 
happen only in the large errors of $\chi^2$; the plot with 
$\Delta \chi^2=2.3$ has almost no change. Therefore, to simplify the 
numerical analysis and to reduce the scanned parameters,  it is 
reasonable in this study to assume $\lambda_{6,7}\ll 1$. Since the 
influence of $|\chi_F|\leq 1$ should be smaller, to get the typical
 contributions from FCNC effects, we illustrate our studies by 
setting  $\chi_F = \pm 1$ in the whole numerical analysis.

With $\lambda_{6,7}\ll 1$, we present the allowed regions for 
$\sin\alpha$ and $\tan\beta$ in Fig.~\ref{fig:satb}, where the left, middle and right panels stand for the 2HDM type-II,  type-III with $\chi_F= -1$ and type-III with $\chi_F = +1$, respectively, and in each plot we show the constraints at $68\%$ CL (green), 95.5$\%$ CL (red) and 99.7$\%$ CL (black). Our results in type-II are consistent with those obtained by the authors in Refs.~\cite{Celis:2013rcs,Ferreira:2011aa} when the same conditions 
 are chosen.  By the plots, we see that in type-III with $\chi_F=-1$,  , due
 to the sign of coupling being  the same as type-II, the allowed values for
 $\sin\alpha$ and $\tan\beta$ are further shrunk; especially $\sin\alpha$ is
 limited to be less than $0.1$. On the contrary, type-III with $\chi_F=+1$,
 the allowed values of $\sin\alpha$ and $\tan\beta$ are broad. 

As discussed before, the decoupling limit occurs at $\alpha\to \beta -\pi/2$, i.e. $\sin\alpha=-\cos\beta< 0$. Since we regard the masses of new scalars as free parameters and scan them in the regions shown in Eq.~(\ref{numbers}), therefore, the three plots in Fig.~\ref{fig:satb} cover lower and heavier mass of charged Higgs. We further check that  $\sin\alpha >0$ could be excluded at 95.5(99.7)$\%$ CL when $m_{H^\pm} \geq 585(690)$ GeV. The main differences between type-II and type-III are the Yukawa couplings as shown in Eq.~(\ref{eq:LhY}). 

\begin{figure}[hptb]\centering
\includegraphics[width=0.32\textwidth]{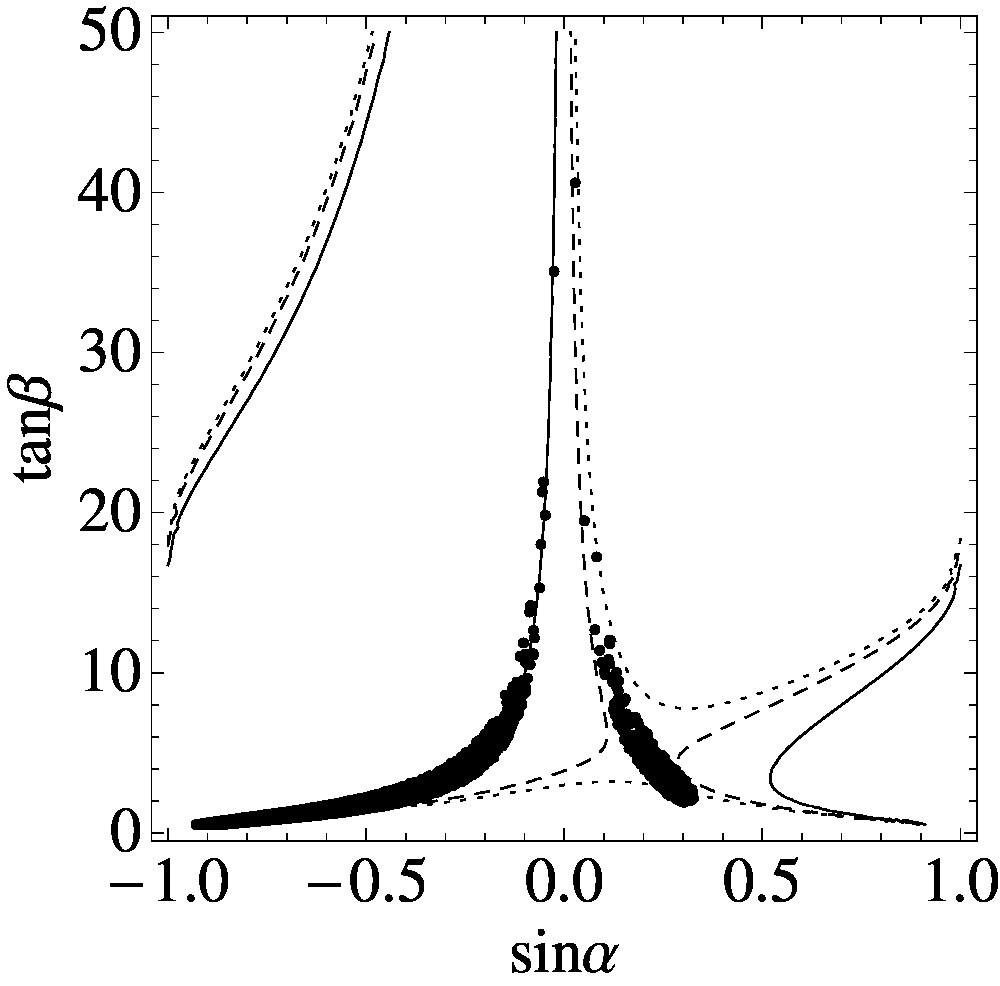}
\includegraphics[width=0.32\textwidth]{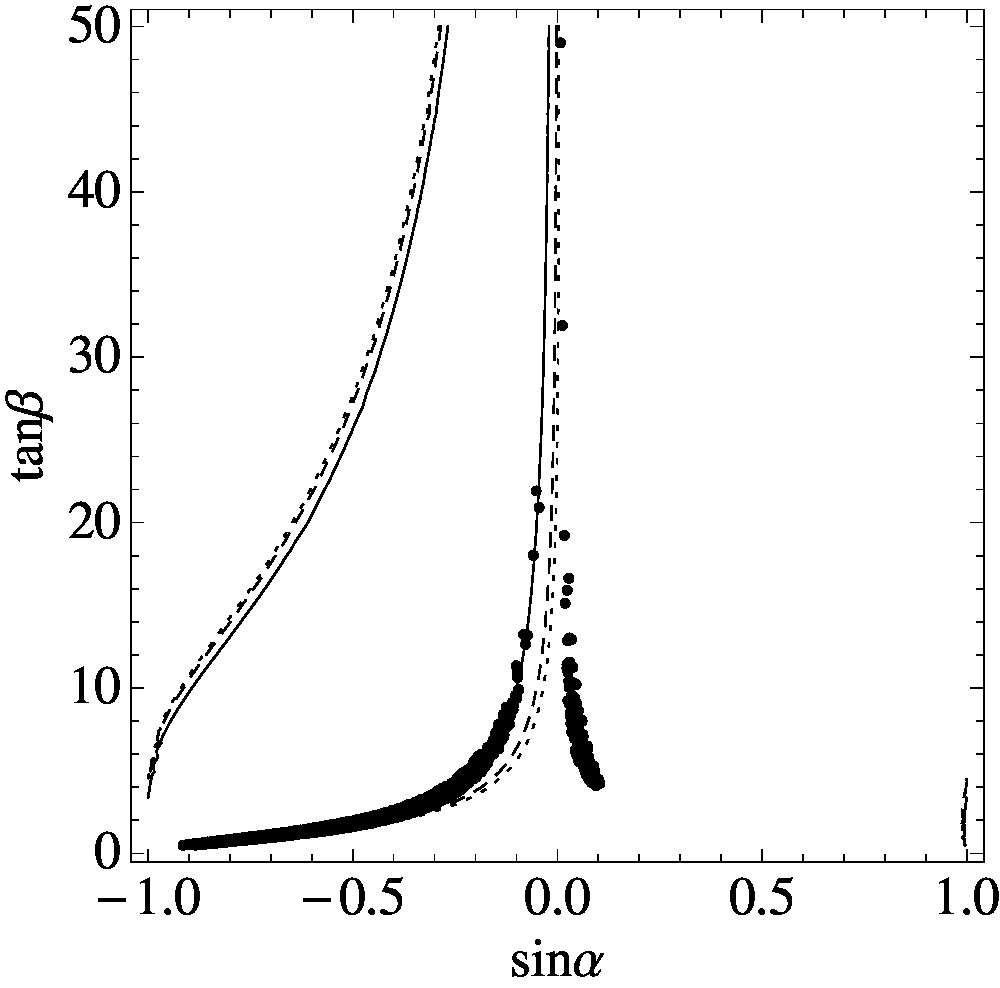}
\includegraphics[width=0.32\textwidth]{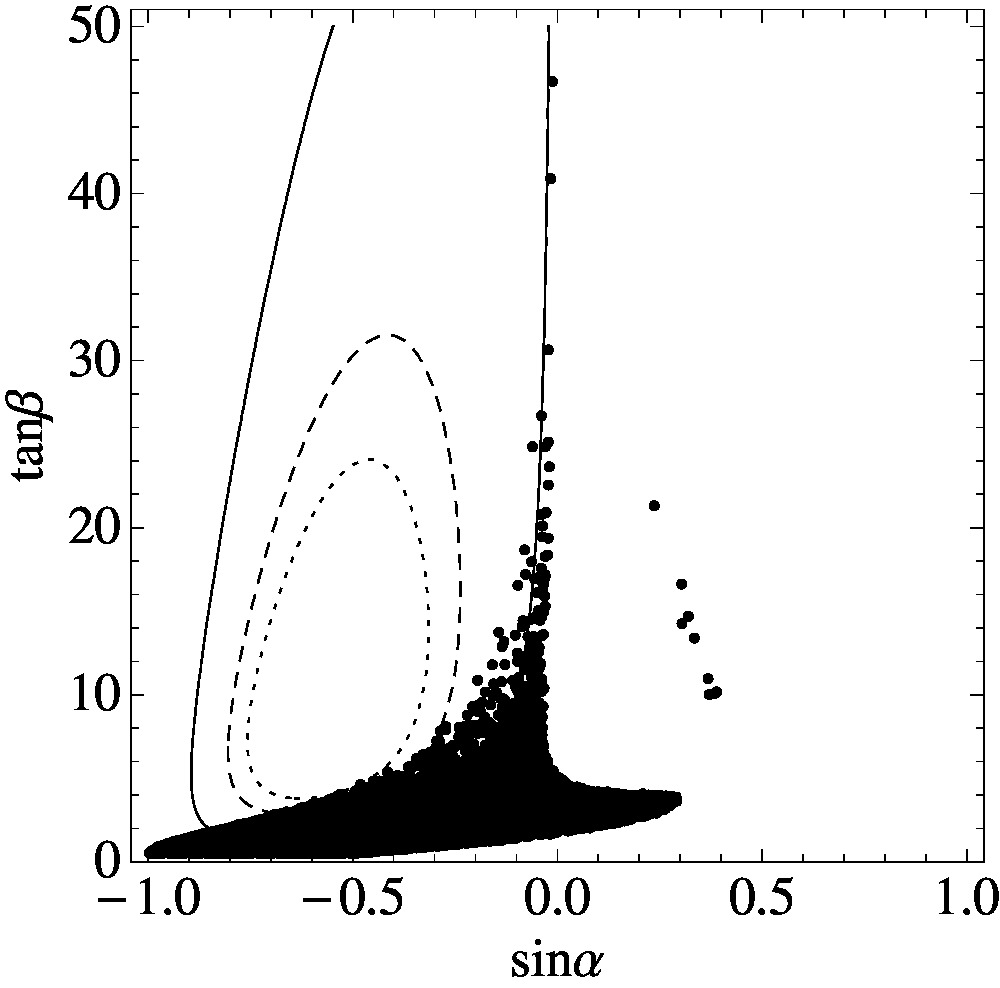}
\caption{$\kappa_{g}$ as a function of  $\sin\alpha$ and $\tan\beta$ in type-II (left) and type-III with $\chi_F=(-1, +1)$ (middle, right), where 
 the solid, dashed and dotted lines in each plot stand for the decoupling limit (DL) of SM, $15\%$ deviation from DL and $20\%$ deviation from DL, respectively. The dotted points are the allowed values of parameters resulted from Fig.~\ref{fig:satb}. }
\label{checkpl}
\end{figure}

\begin{figure}[hptb]
\includegraphics[width=0.32\textwidth]{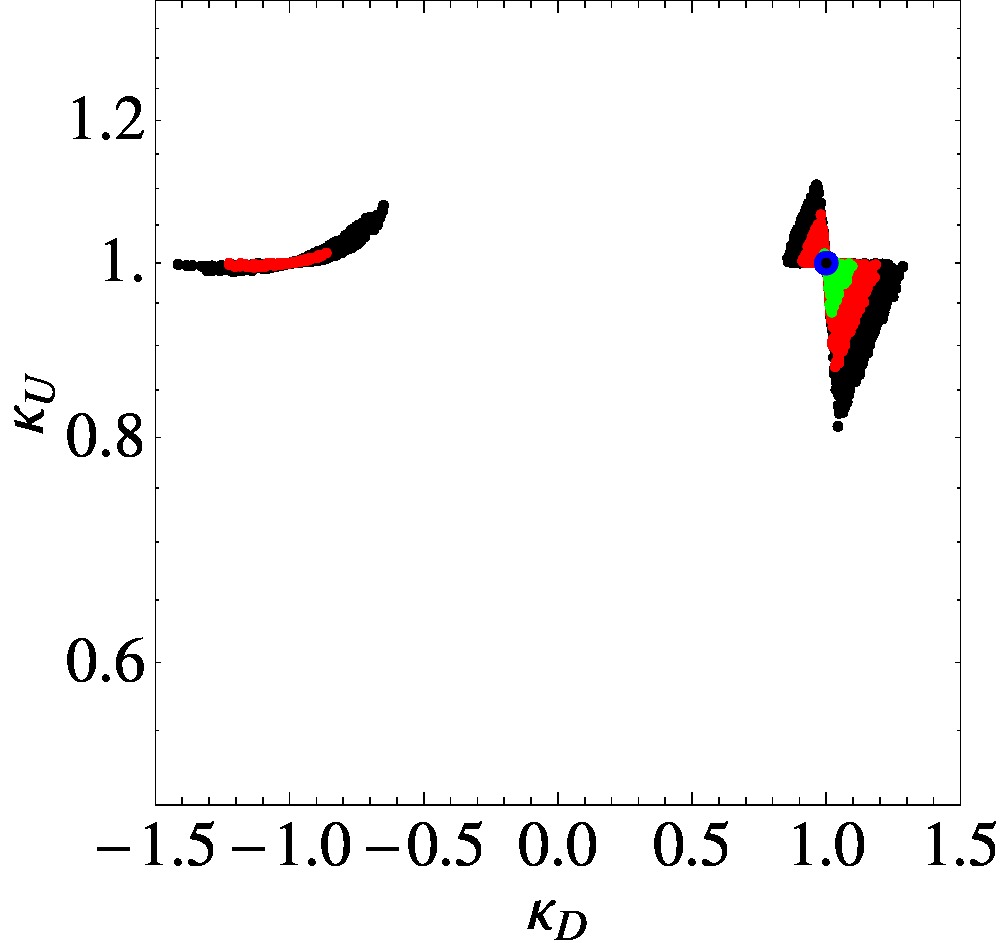}
\includegraphics[width=0.32\textwidth]{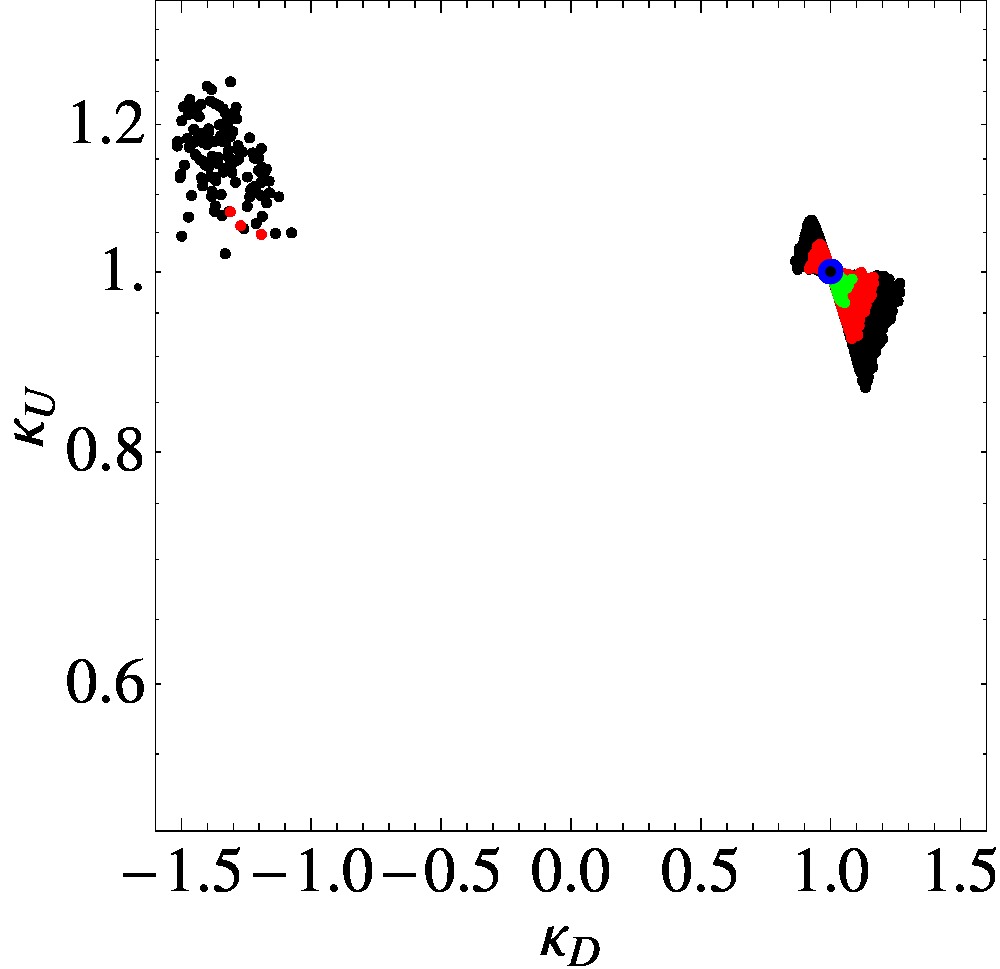}
\includegraphics[width=0.32\textwidth]{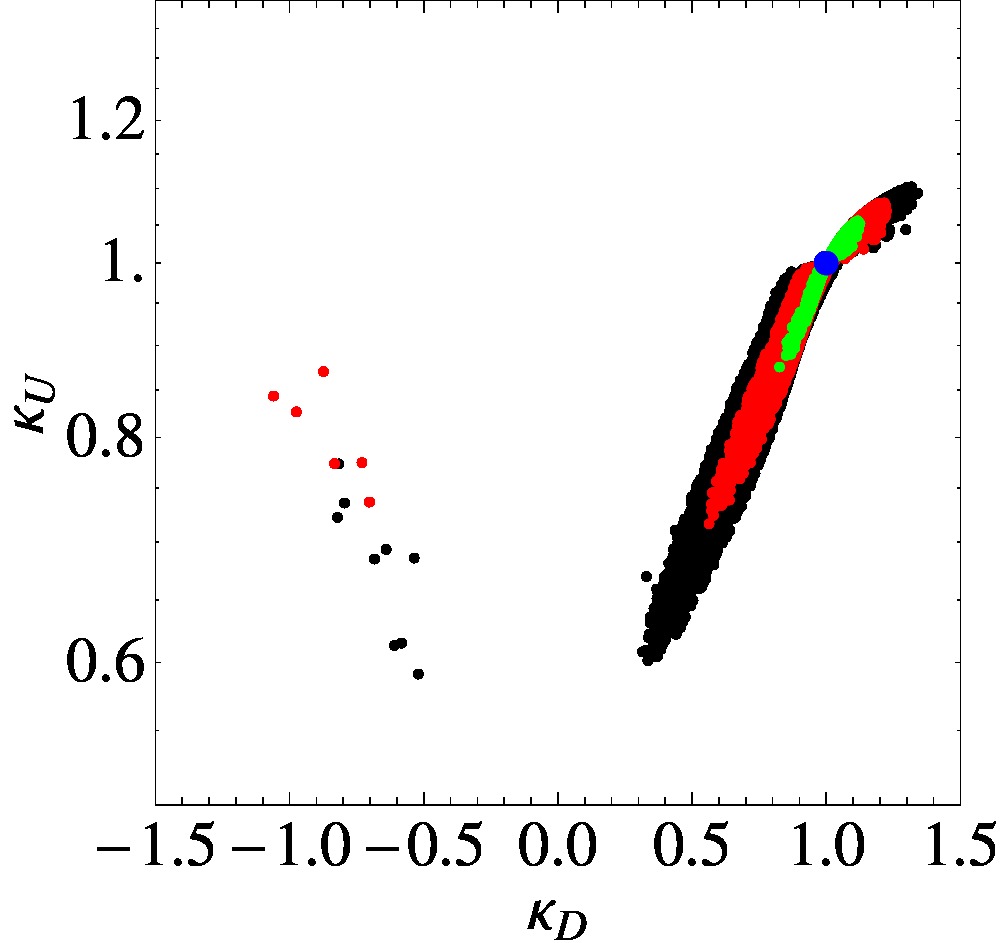}
\caption{
Correlation of $\kappa_D$ and $\kappa_U$, where the left, middle and right panels represent the allowed values in type-II,  type-III with $\chi_F=-1$ and  type-III with $\chi_F=+1$, respectively and  the results of Fig.~\ref{fig:satb} are applied.  }
\label{fig:cdcu}
\end{figure}

\begin{figure}[hptb]
\includegraphics[width=0.32\textwidth]{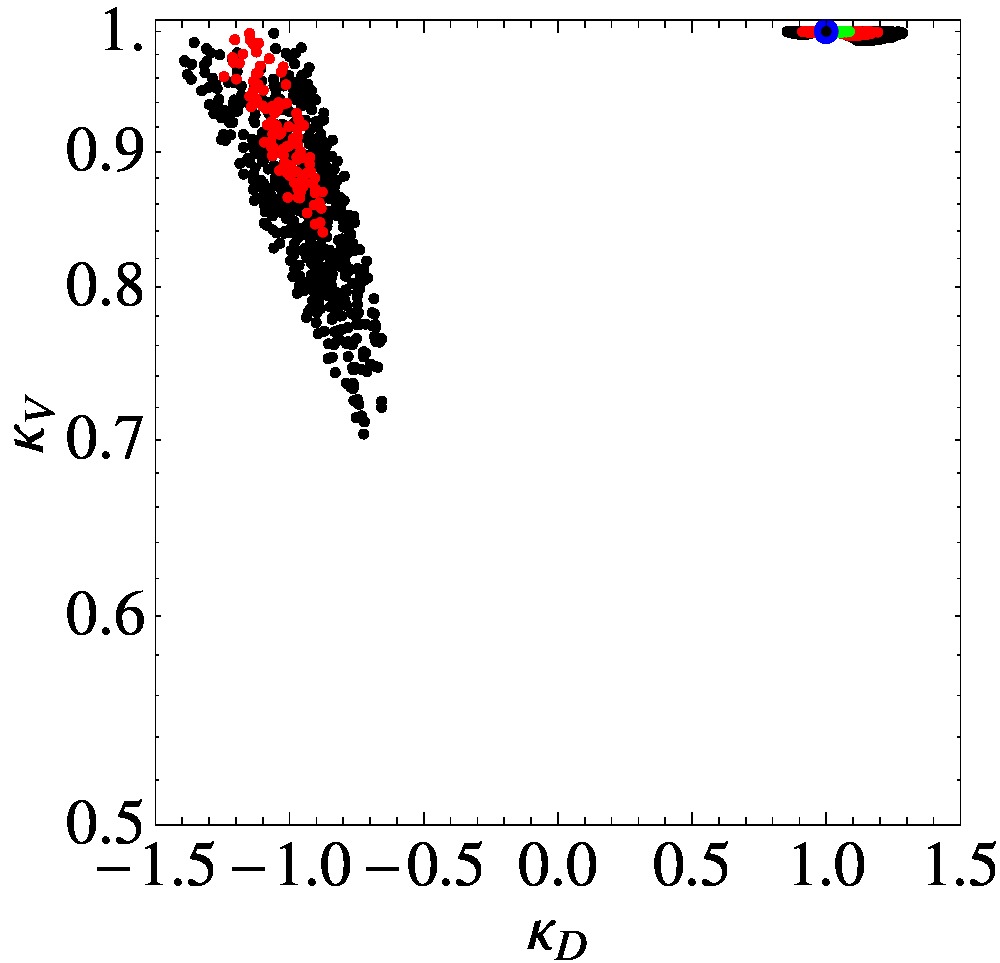}
\includegraphics[width=0.32\textwidth]{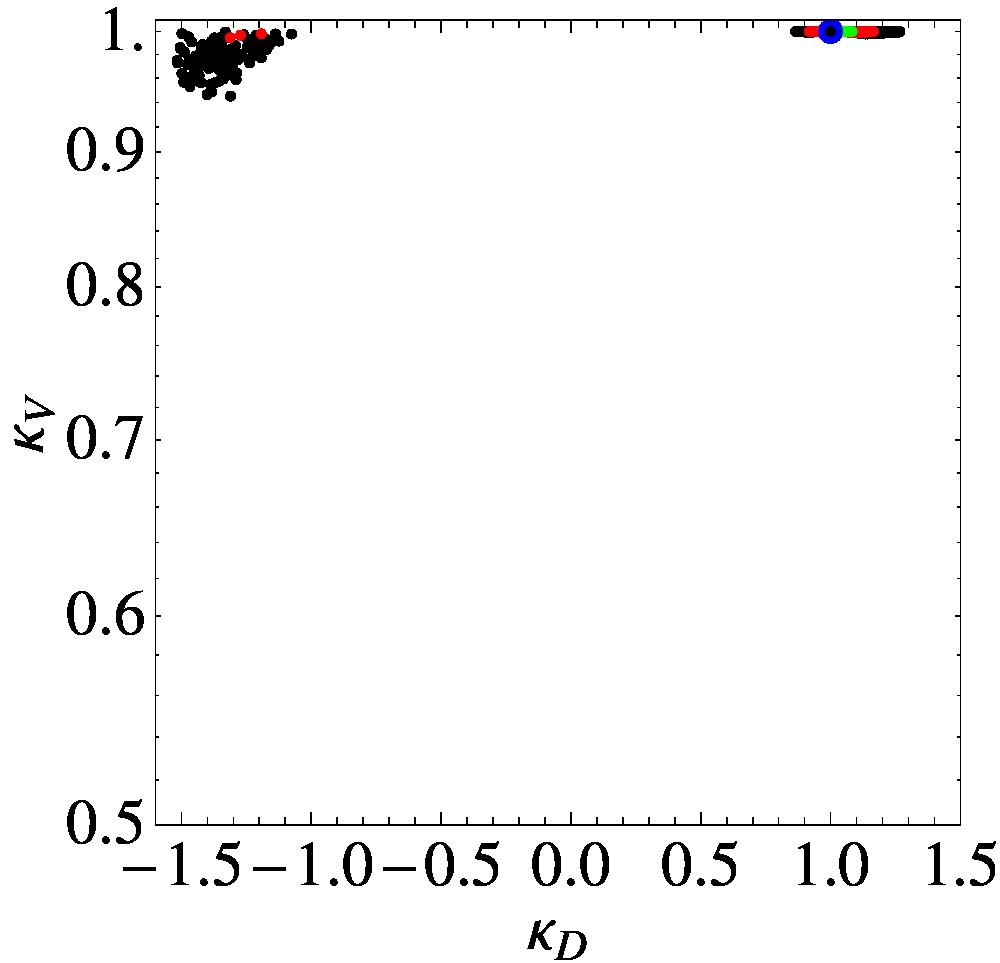}
\includegraphics[width=0.32\textwidth]{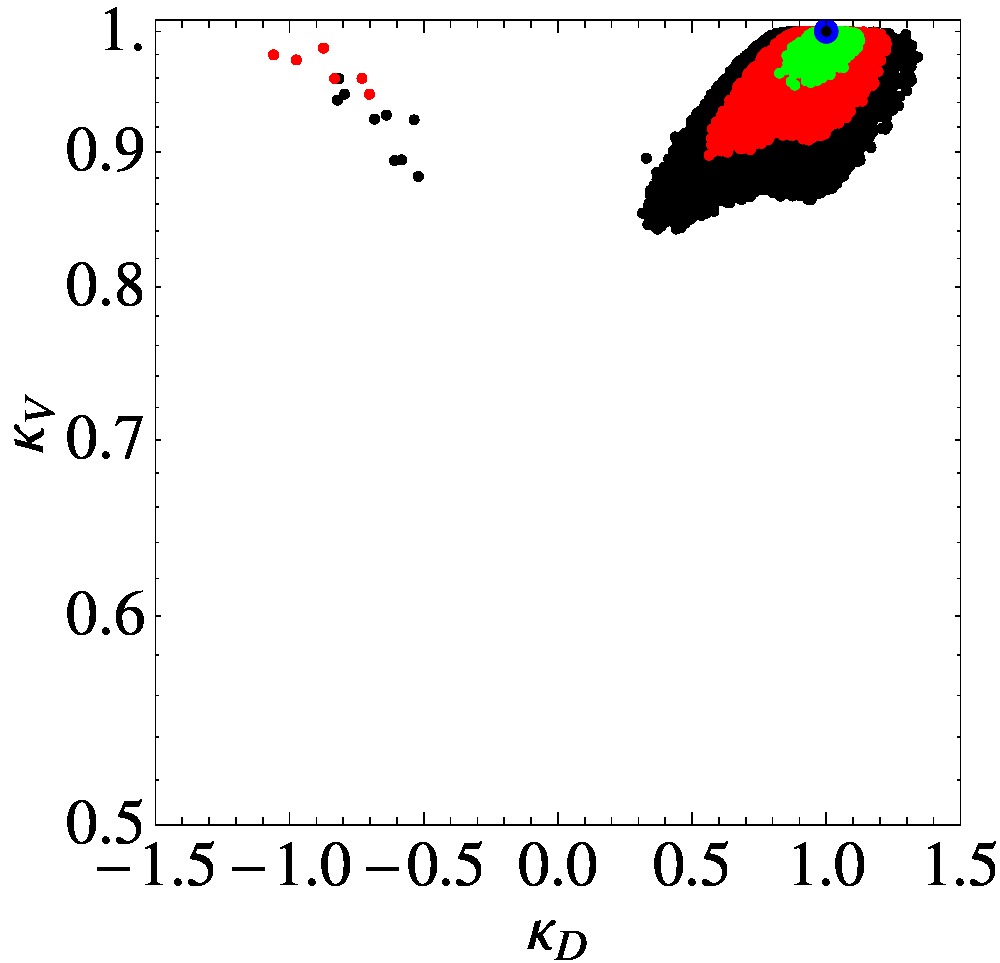}
\caption{
The legend is the same as that in Fig.~\ref{fig:cdcu}, but for the correlation of $\kappa_V$ and $\kappa_D$.  }
\label{fig:cdcv}
\end{figure}

In order to see the influence of  the new effects of type-III, we plot the allowed $\kappa_g$ as  a function  of $\sin\alpha$ and $\tan\beta$ in Fig.~\ref{checkpl}, where the three plots from left to right correspond to type-II, type-III with $\chi_F=-1$ and type-III with $\chi_F=+1$, the solid, dashed  and dotted lines in each plot stand for the decoupling limit (DL) of SM,   $15\%$ deviation from DL and $20\%$ deviation from DL, respectively.  For comparisons, we also put the results of $99.7\%$ in Fig.~\ref{fig:satb} in each plot. By the analysis, we see that the deviations of $\kappa_g$ from DL  in $\chi_F=+1$ are clear and significant while the influence of $\chi_F=-1$ is small. 
It is pointed out that  a wrong sign Yukawa coupling to down type quarks could happen in type-II 2HDM \cite{wrong2,Gunion:2002zf}. For understanding the sign flip, we rewrite the $\kappa_D$ defined in Eq.~(\ref{scaling3}) to be
\be
\kappa_{D}= -\frac{\sin\alpha}{\cos\beta} \left( 1- \frac{\chi_F \sin\beta}{\sqrt{2}}\right) + \frac{\chi_F \cos\alpha}{\sqrt{2}}\,.
\label{eq:kD}
\ed 
In the type-II case, we know that  in the decoupling limit $\kappa_D=1$, but $\kappa_D < 0$ if $\sin\alpha >0$.  According to the results in the left panel of Fig.~\ref{fig:satb}, $\sin\alpha >0$ is allowed when the errors of best fit are taken by $2\sigma$ or 3$\sigma$. The situation in type-III is more complicated. From  Eq.~(\ref{eq:kD}), we see that the factor in the brackets is always positive, therefore, the sign of the first term should be the same as that in type-II case. However, due to $\alpha \in [ -\pi/2, \pi/2]$, the sign of the second term in Eq.~(\ref{eq:kD}) depends on the sign of $\chi_F$. For $\chi_F = -1$, even $\sin\alpha < 0$, $\kappa_D$ could be negative when the first term is smaller than the second term. For $\chi_F=+1$, if $\sin\alpha >0$ and the first term is over the second term, $\kappa_D < 0$ is still allowed.  In order to understand the available values of $\kappa_D$ when the constraints are considered, we present the correlation of $\kappa_U$ and $\kappa_D$ in Fig.~\ref{fig:cdcu}, where the panels from left to right stand for type-II, type-III with $\chi_F = -1$ and type-III with $\chi_F=+1$. In each plot, the results obtained by $\chi$-square fitting are applied. The similar correlation of $\kappa_V$ and $\kappa_D$ is presented in Fig.~\ref{fig:cdcv}. By these results, we find that comparing with type-II model,  the negative $\kappa_D$ gets more strict limit in type-III, although a wider  parameter space for $\sin\alpha > 0$ is allowed in type-III with $\chi_F = +1$. 
%

\begin{figure}[hptb]
\includegraphics[width=0.32\textwidth]{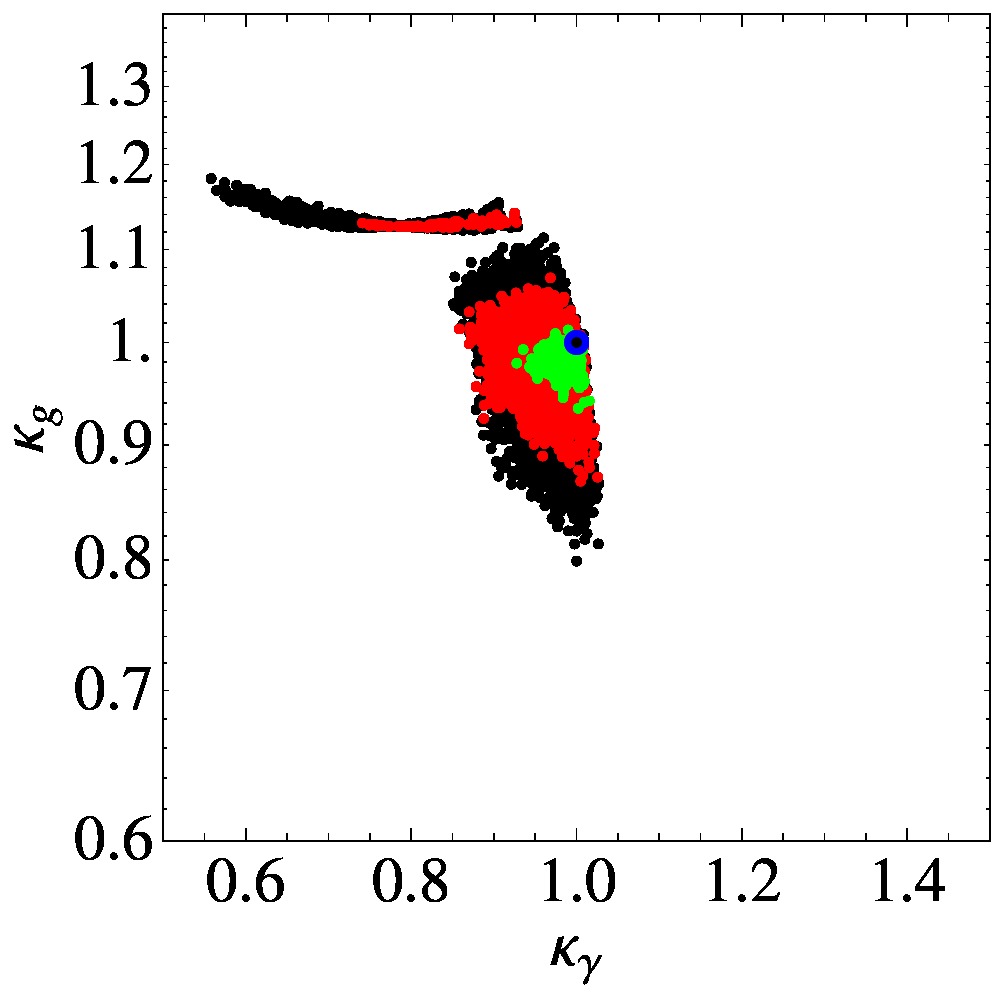}
\includegraphics[width=0.32\textwidth]{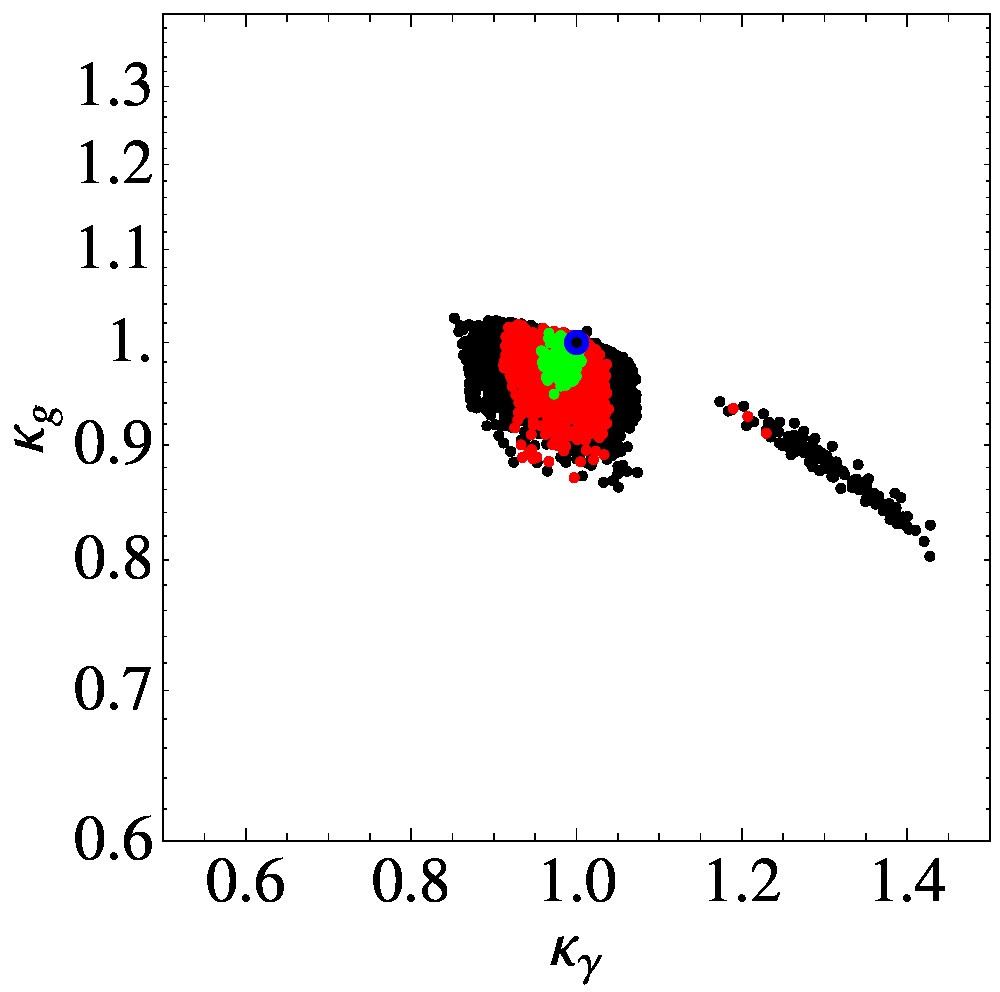}
\includegraphics[width=0.32\textwidth]{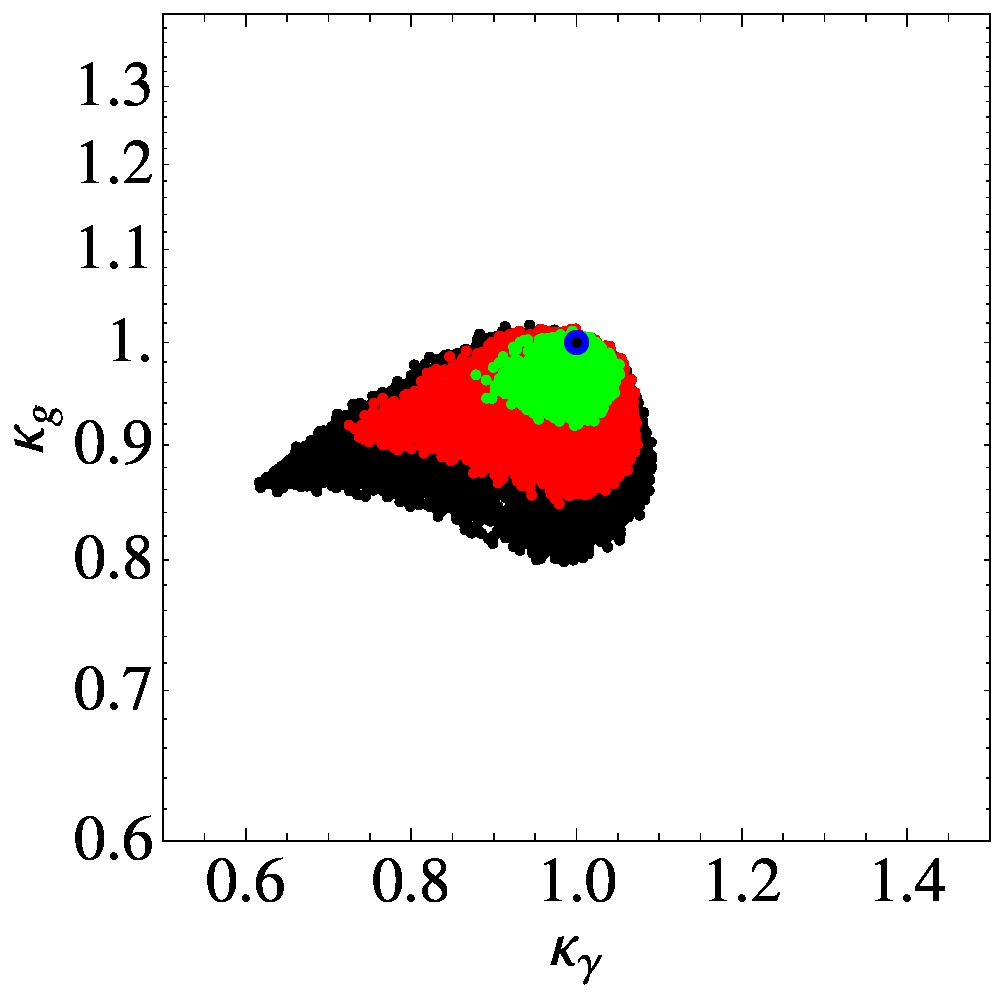}
\caption{Correlation of $\kappa_\gamma$ and $\kappa_g$, where the left, middle and right panels represent the allowed values in type-II,  type-III with $\chi_F=-1$ and  type-III with $\chi_F=+1$, respectively and  the results in Fig.~\ref{fig:satb} obtained by $\chi$-square fitting  are applied.}
\label{fig:gaga}
\end{figure}

Besides the scaling factors of tree level Higgs decays, $\kappa_{D,U}$ and $\kappa_V$, it is also interesting to understand the allowed values for loop induced processes in 2HDM, e.g. $h\to \gamma\gamma$, $gg$, and  $Z\gamma$, etc. It is known that the differences in the associated couplings between $h\to \gamma \gamma$ and $gg$  are the colorless $W$-, $\tau$-, and $H^\pm$-loop. By Eq.~(\ref{scaling2}), we see that the contributions of  $\tau$ and $H^\pm$ are small, therefore, the main difference is from the $W$-loop in which the $\kappa_V$ involves. By using the $\chi$-square fitting approach and with the inputs of  the experimental data and theoretical constraints, the allowed regions of $\kappa_\gamma$ and $\kappa_g$ in type-II and type-III are displayed in Fig.~\ref{fig:gaga}, where the panels from left to right are type-II, type-III with $\chi_F=-1$ and $+1$; the green, red and black colors in each plot stand for the $68\%$, $95.5\%$ and $99.7\%$ CL, respectively. We find that except  a slightly lower $\kappa_\gamma$ is allowed in type-II, the first two plots have similar results. The situation can be understood from Figs.~\ref{fig:cdcu} and \ref{fig:cdcv}, where the $\kappa_U$ in both models is similar while $\kappa_V$ in type-II could be smaller in the region of negative $\kappa_D$; that is,  a smaller $\kappa_V$ will lead a smaller $\kappa_\gamma$. In $\chi_F=+1$ case, the allowed values of $\kappa_\gamma$ and $\kappa_g$ are localized in a wider region. 
%

\begin{figure}[h!]
\includegraphics[width=0.32\textwidth]{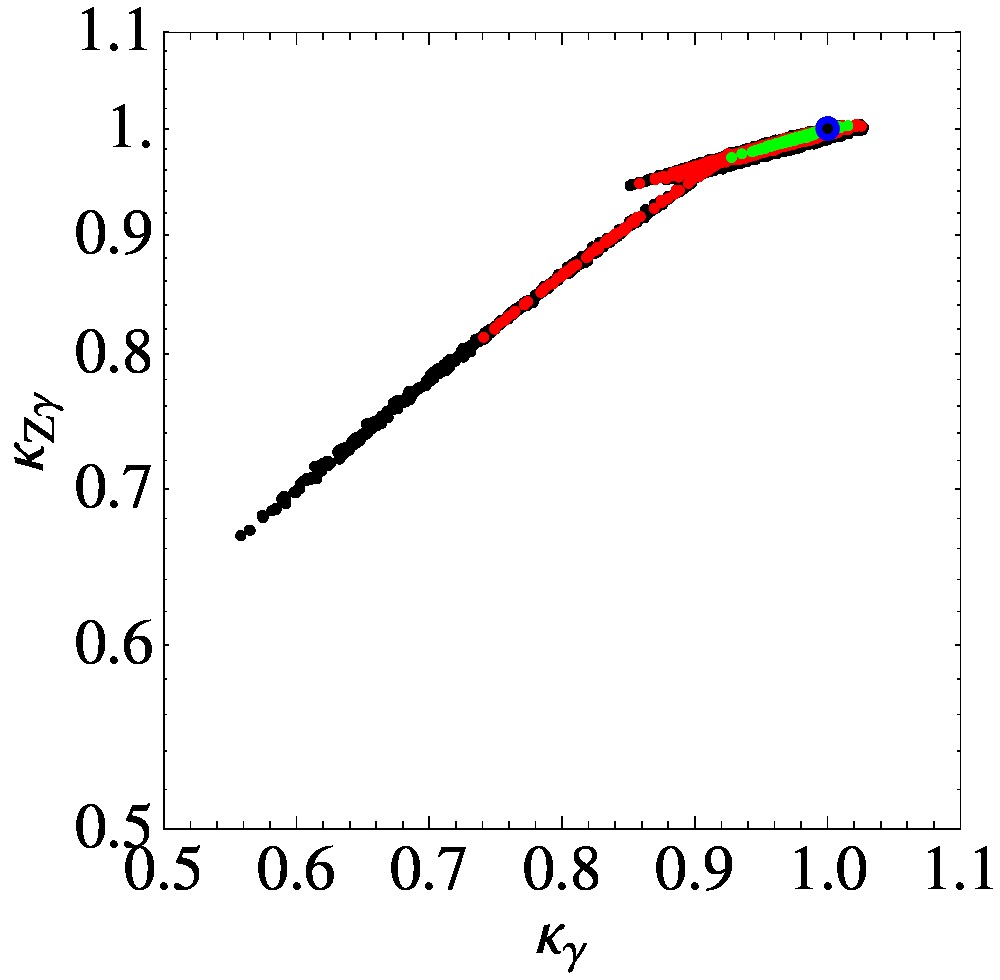}
\includegraphics[width=0.32\textwidth]{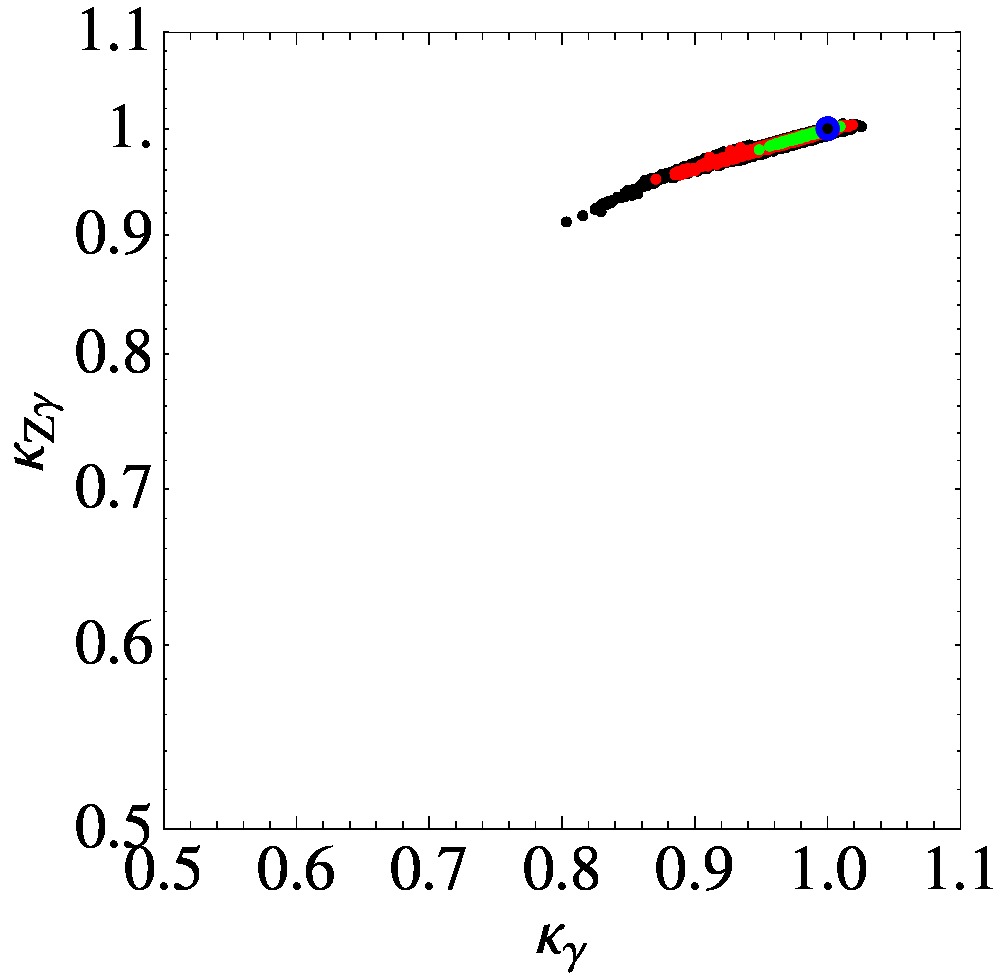}
\includegraphics[width=0.32\textwidth]{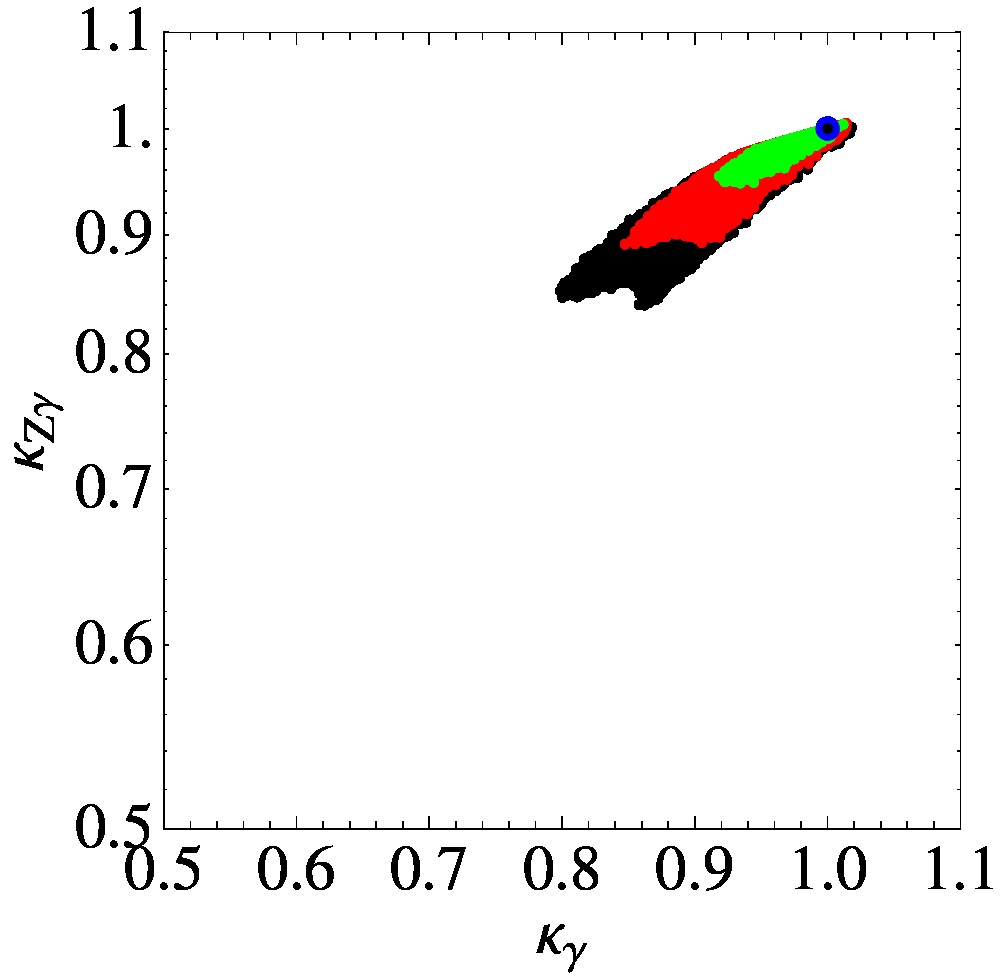}
\caption{The allowed regions in $(\kappa_{\gamma}, \kappa_{Z\gamma})$ plan 
after imposing theoretical and experimental constrains. 
Color coding the same as Fig.~\ref{fig:satb}}
\label{fig:zga}
\end{figure}

\begin{figure}[h!]
\includegraphics[width=0.32\textwidth]{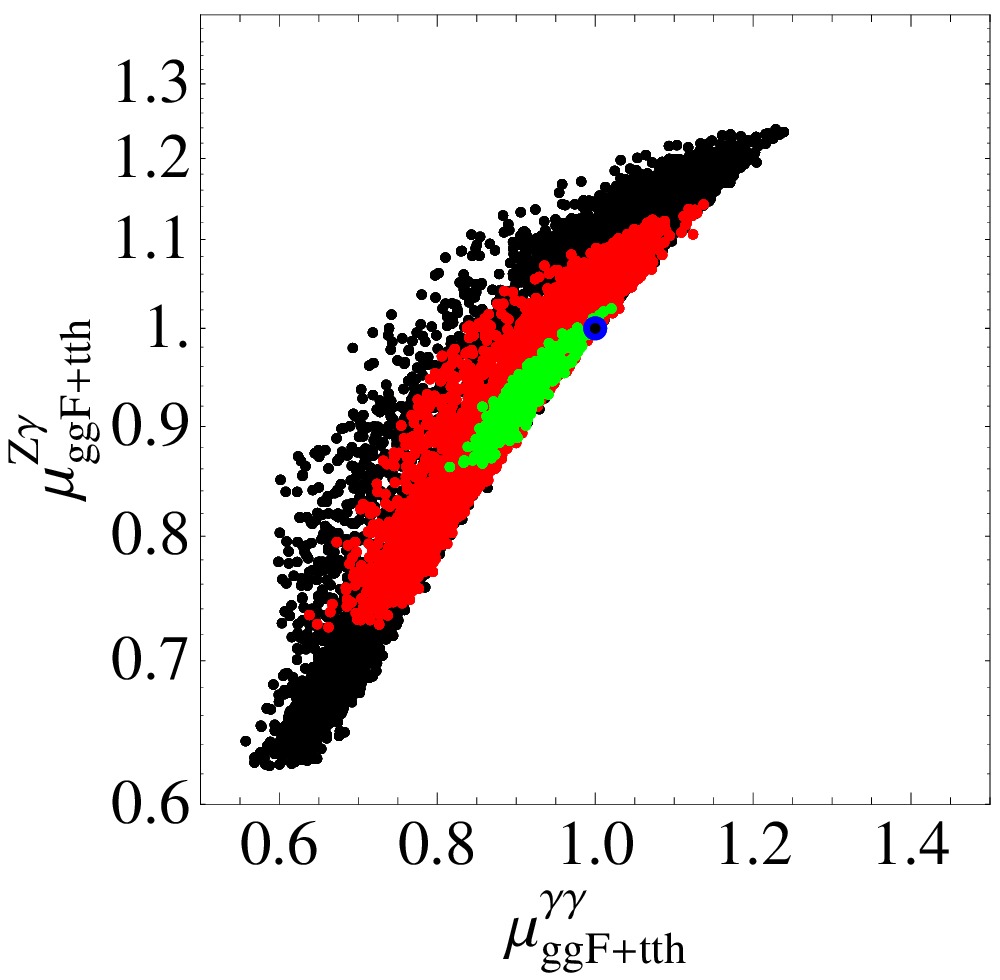}
\includegraphics[width=0.32\textwidth]{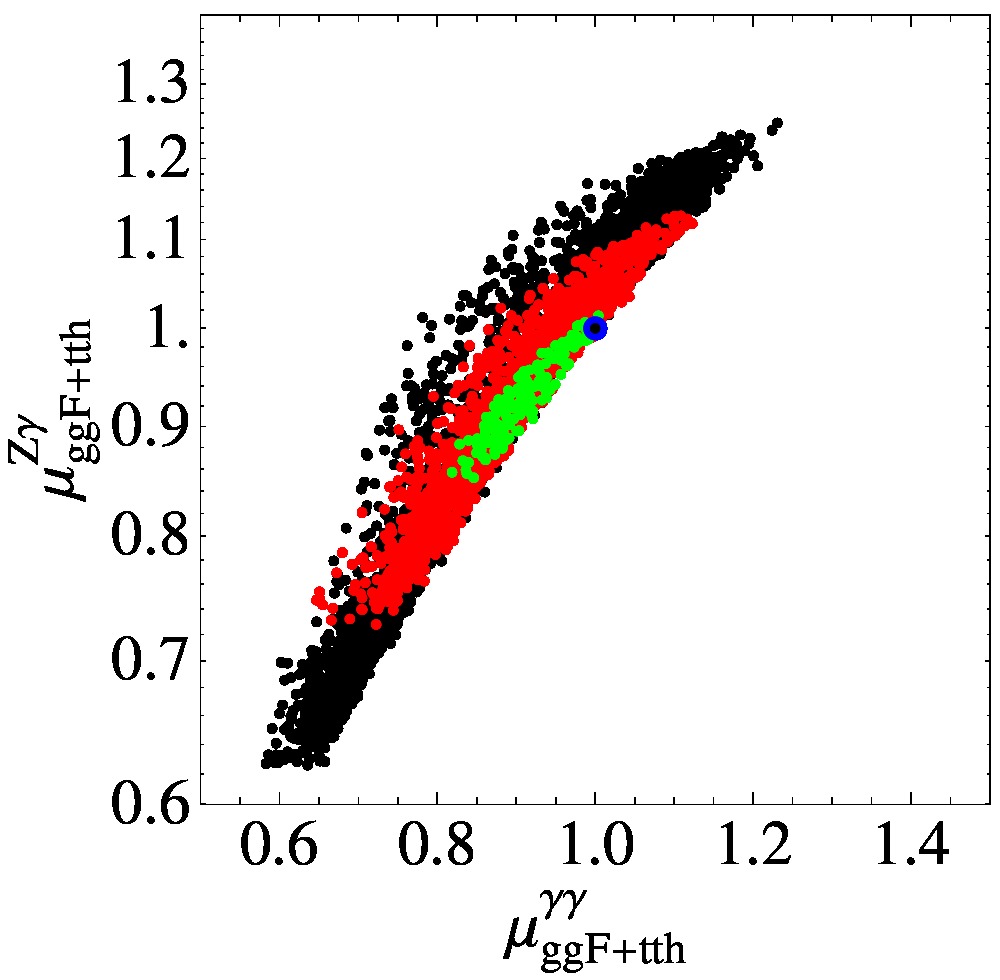}
\includegraphics[width=0.32\textwidth]{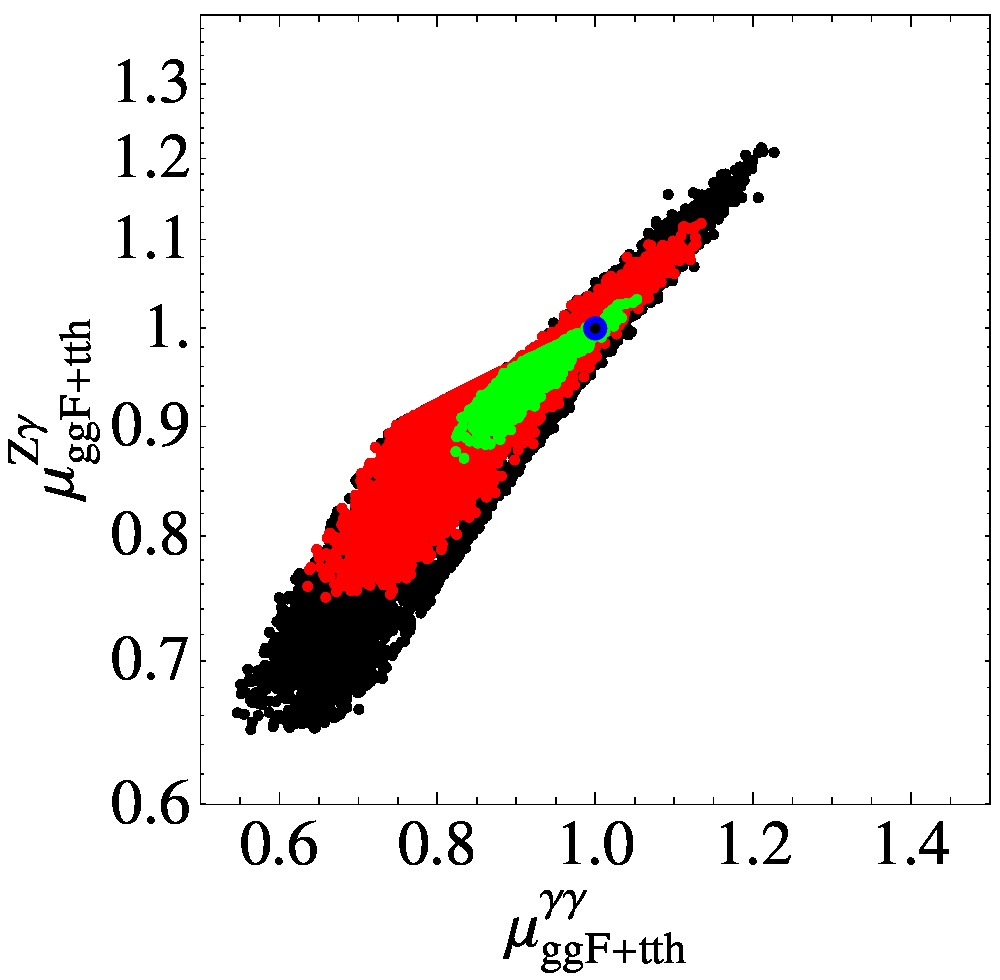}
\caption{Correlation between $\mu^{\gamma\gamma}_{ggF+tth}$ and $\mu^{Z\gamma}_{ggF+tth}$
 at $\sqrt{s} = 13$ TeV after imposing theoretical and experimental
 constrains. Left, middle and right panels represent the allowed values in type-II,  type-III with $\chi_F=-1$ and  type-III with $\chi_F=+1$, respectively and  the results in Fig.~\ref{fig:satb} obtained by $\chi$-square fitting  are applied.}
\label{fig:zga13}
\end{figure}

It is known that except the different gauge couplings, the loop diagrams for
$h\to Z\gamma$ and $h\to \gamma\gamma$ are exact the same. One can understand the
loop effects by the numerical form of Eq.~(\ref{scaling2}). Therefore, we expect
the correlation between $\kappa_{Z\gamma}$ and $\kappa_\gamma$ should behave
like a linear relation. We present the correlation between $\kappa_\gamma$ and
$\kappa_{Z\gamma}$ in Fig.~\ref{fig:zga}, where the legend is the same as that
for Fig.~\ref{fig:gaga}.  From the plots, we see that in most region
$\kappa_{Z\gamma}$ is less than the SM prediction. The type-III with
$\chi_F=-1$ gets more strict constraint and the change is within $10\%$. For
$\chi_F=+1$, the deviation of $\kappa_{Z\gamma}$ from unity could be over
$10\%$. From run I data, the LHC has an upper bounds on 
$h\to Z\gamma$, at run II this decay mode will be probed.
We give the predictions at 13 TeV LHC  for the signal strength 
$\mu^{\gamma\gamma}_{ggF+tth}$  and $\mu^{Z\gamma}_{ggF+tth}$ in
Fig.~\ref{fig:zga13}. Hence, with the theoretical and experimental
constraints, $\mu^{Z\gamma}_{ggF+tth}$ is bounded and could be
$\cal{O}$(10\%) away from SM at 68$\%$CL.
%
\section{$t\to ch$ decay}

In this section, we study the flavor changing $t\to c h$ process 
in type-III model.
Experimentally, there have been intensive activities to 
explore the top FCNCs.  CDF, D0 and LEPII collaborations have 
reported some bounds on top FCNCs. At the LHC with rather 
large top cross section, ATLAS and CMS search for top FCNCs and
put a limit on the branching fraction which is $Br(t\to ch)< 0.82$ \% for ATLAS \cite{ATLAStch} and  $Br(t\to ch)< 0.56$ \% for CMS 
\cite{CMStch}. 
Note that CMS limit is slightly better than ATLAS limit. 
CMS search for $t\to ch$ in different channels: $h\to \gamma\gamma$, $WW^*$, $ZZ^*$, and  $\tau^+\tau^-$ while ATLAS used only diphoton channel. With the high luminosity option of the LHC, the above limit  will be improved to reach about $Br(t\to ch)< 1.5 \times 10^{-4}$  \cite{ATLAStch} for ATLAS detector.

From the Yukawa couplings in  Eq,~(\ref{eq:LhY}), the partial width for $t\to ch$ decay is given by
\be
\Gamma(t\to ch) &=&
\left(\frac{\cos(\beta-\alpha)X_{23}^u}{\sin\beta}\right)^2\frac{m_t}{32\pi} \left( (x_c + 1)^2 - x^2_h\right) \non \\
&\times & \sqrt{1-(x_h - x_c)^2}\sqrt{1-(x_h + x_c)^2} 
\ed
where $x_c = m_c/m_t$, $x_h =m_h/m_t$ and $X_{23}^u$ is a free parameter and dictates the FCNC effect. It is clear from
the above expression that the partial width of $t\to ch$ is proportional to
$\cos(\beta-\alpha)$. As seen in the previous section, in the case where $h$
is SM-like, $\cos(\beta-\alpha)$  is constrained by LHC data  to be rather
small and the  $t\to ch$ branching fraction is limited. As we will see later
in 2HDM type-II with flavor conservation the rate for $t\to ch$ is much
smaller than type-III~\cite{thc:Arhrib}.
Since we assume that the charged Higgs is heavier than 400 GeV,
the total decay width of the top contains only $t\to ch$ and $t\to bW$.
With $m_h=125.36$ GeV, $m_t=173.3$ GeV and $m_c=1.42$ GeV, the total width can be written as
\be
\Gamma_t  = \Gamma^{SM}_t +
0.0017 \left(\frac{\cos(\beta-\alpha)X^u_{23}}{\sin\beta}\right)^2\,\,{\rm GeV}
\ed
where $\Gamma^{SM}_t$ is the partial decay rate for $t\to Wb$ and is given by
 \be
 \Gamma^{SM}_t= \frac{G_F m_t^3}{8\pi \sqrt{2}}(1-\frac{m_W^2}{m_t^2})^2
     (1+2 \frac{m_W^2}{m_t^2}) (1-2 \frac{\alpha_s(m_t)}{3 \pi} (2\frac{\pi^2}{3}-\frac{5}{2}))=1.43 \, {\rm GeV}\non
      \ed
in which the QCD corrections have been included. By the above numerical expressions together with the current limit 
 from ATLAS and CMS, the limit on the $tch$ FCNC coupling is found by
\begin{eqnarray}
\left(\frac{\cos(\beta-\alpha)X^u_{23}}{\sin\beta}\right) < 2.2 \quad\quad {\rm for}
\quad \quad Br(t\to ch)< 8.2  \times 10^{-3}\,, \nonumber\\
\left(\frac{\cos(\beta-\alpha)X^u_{23}}{\sin\beta}\right) < 0.36 \quad\quad {\rm for}
\quad \quad Br(t\to ch)< 5.6 \times 10^{-3} \label{eq:bounds}
 \end{eqnarray}
in agreement with~\cite{wshou}.

\begin{figure}[h!]
\begin{center}
\includegraphics[width=0.33\textwidth]{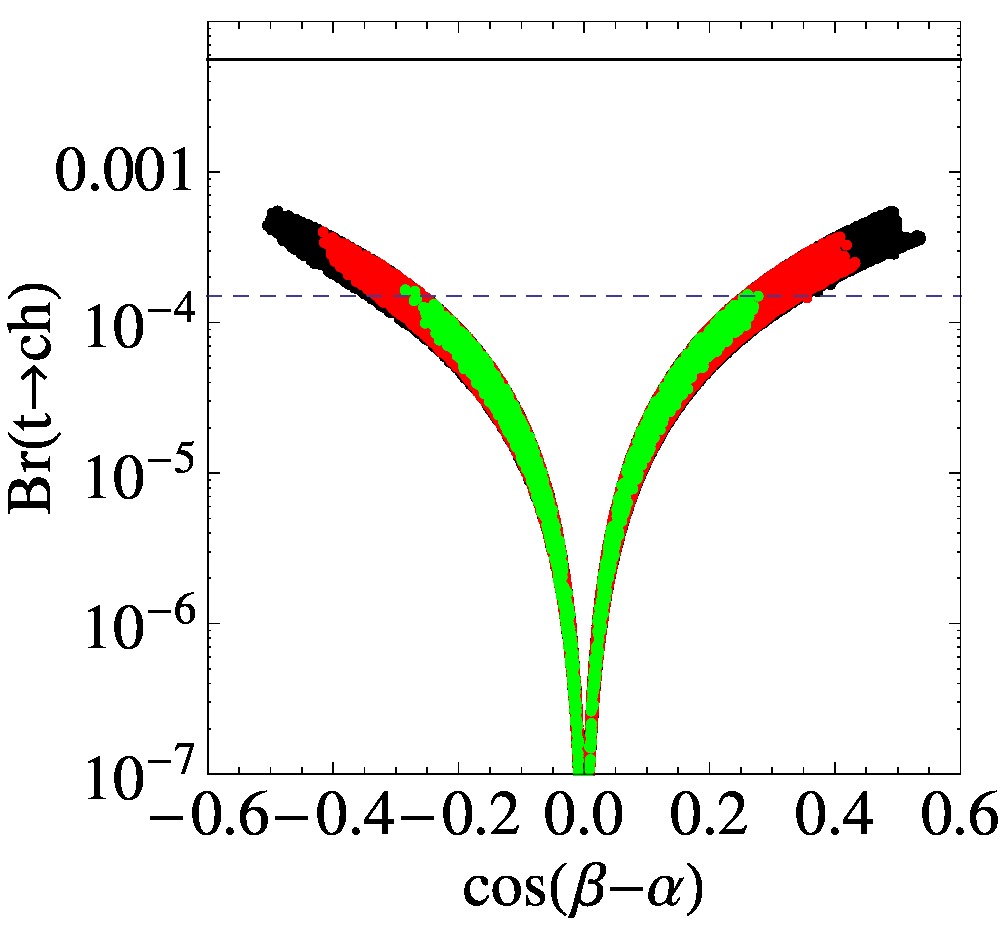}
\includegraphics[width=0.32\textwidth]{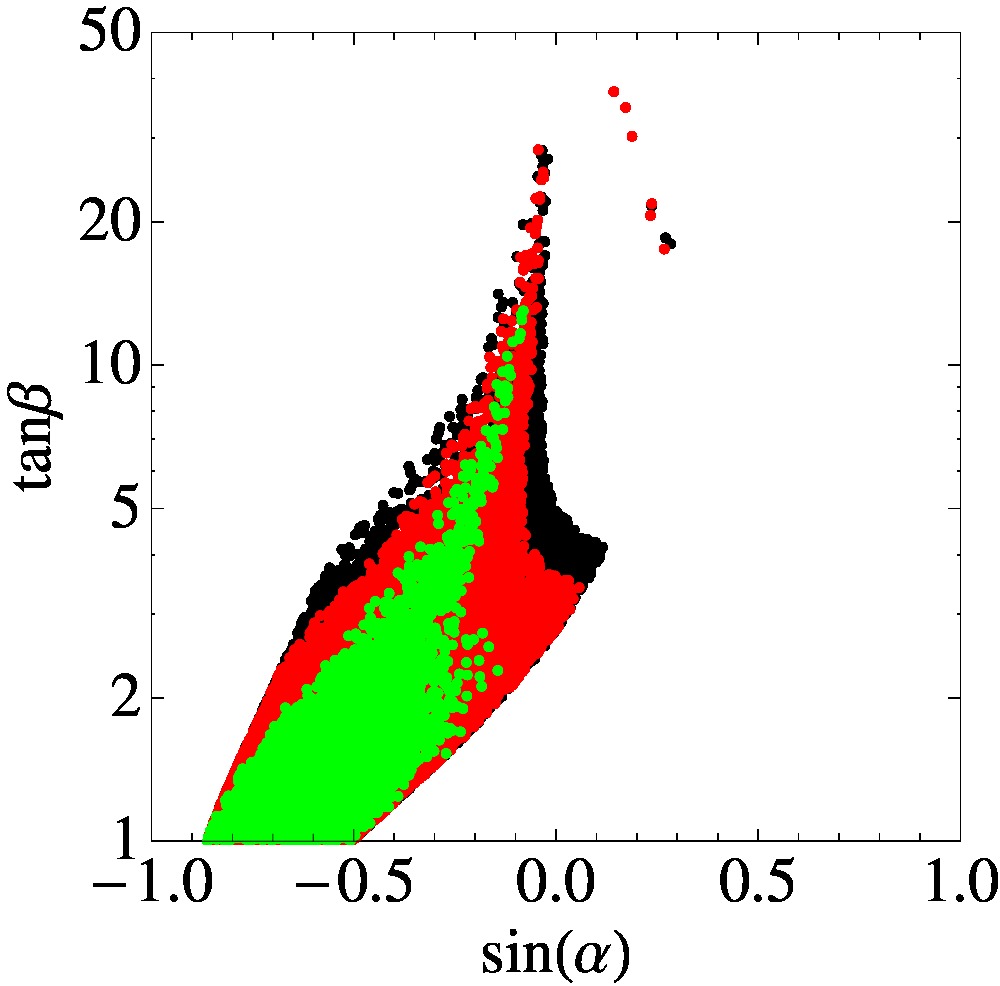}
\end{center}
\caption{Left: Branching ratio of Br$(t\to ch)$ as a function of
    $\cos(\beta -\alpha)$, the two horizontal lines correspond to 
 LHC actual limit (upper line) and expected limit from ATLAS with 
3000 fb$^{-1}$ luminosity (dashed line). Right panel: allowed parameters space
in type III with the ATLAS expected limit on $Br(t\to ch)< 1.5 \times 10^{-4}$. }
\label{fig:brtch}
\end{figure}

We perform a systematic scan over 2HDM parameters, as depicted in Eq.~(\ref{numbers}), taking into account LHC and theoretical constraints.
 Although $X^u_{23}$ is a free parameter,  in order to suppress the FCNC effects naturally, as stated earlier we adopt $X^u_{23} = \sqrt{m_t m_c}/v \chi^u_{23}$. Since the current experimental measurements only give a upper limit on $t\to h c$, basically $\chi^u_{23}$ is limited by Eq.~(\ref{eq:bounds}) and could be as large as ${\cal O}(1-10^{2})$, depending on the allowed value of $\cos(\beta-\alpha)$. In order to use the constrained results which are obtained from the Higgs measurements and the self-consistent parametrisation $X^u_{33}=m_t /v \chi_F$ which was used before, we assume $\chi^u_{23} = \chi_F = \pm 1$. In our numerical analysis, the results under the assumption should be conservative. 
In Fig.~\ref{fig:brtch}(left) we illustrate the branching fraction of $t\to ch$ in 2HDM-III as a function of $\cos(\beta-\alpha)$.  
The LHC constraints within 1$\sigma$ restrict $\cos(\beta-\alpha)$ to be in the 
range $[-0.27,0.27]$. The branching fraction for $t\to ch$ is slightly above $10^{-4}$ .
The actual CMS and ATLAS constraint on $Br(t\to ch)<5.6 \times 10^{-3}$  does not restrict $ \cos(\beta-\alpha)$.
The expected limit from ATLAS detector with high luminosity 3000 fb$^{-1}$ is depicted as the dashed horizontal line. As one can see, 
the expected ATLAS limit is somehow similar to LHC constraints  within 1$\sigma$.
In the right panel, we show the allowed parameters space in $(\sin\alpha, \tan\beta)$ plan where we apply 
ATLAS expected limit $Br(t\to ch)< 1.5 \times 10^{-4}$.  This plot should be compared to the right panel of Fig.\ref{fig:satb}.
It is then clear that this additional constraint only act on the  3$\sigma$ allowed region from LHC data.

\begin{figure}[h!]
\includegraphics[width=4.8cm,height=6cm]{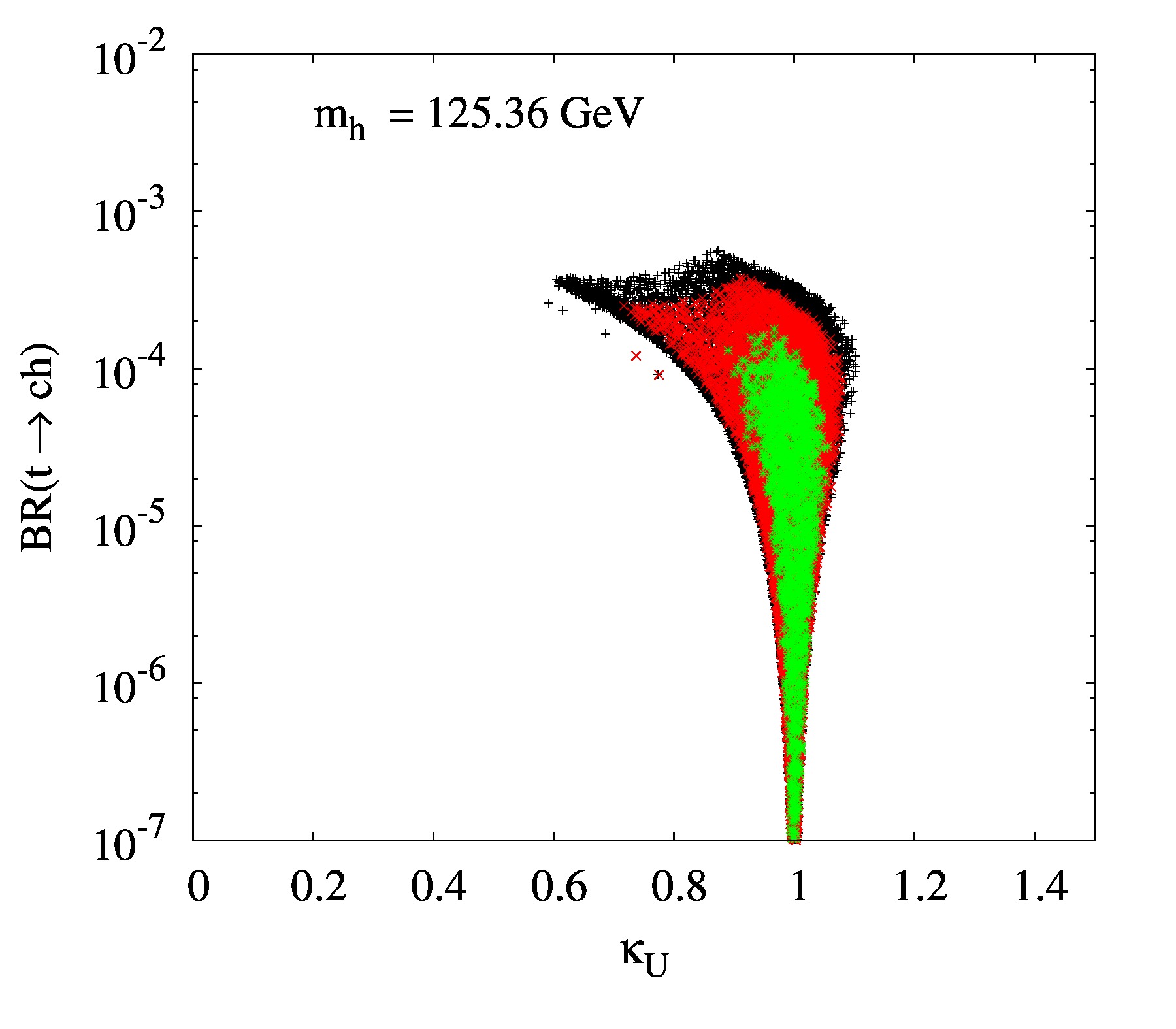}
\includegraphics[width=4.8cm,height=6cm]{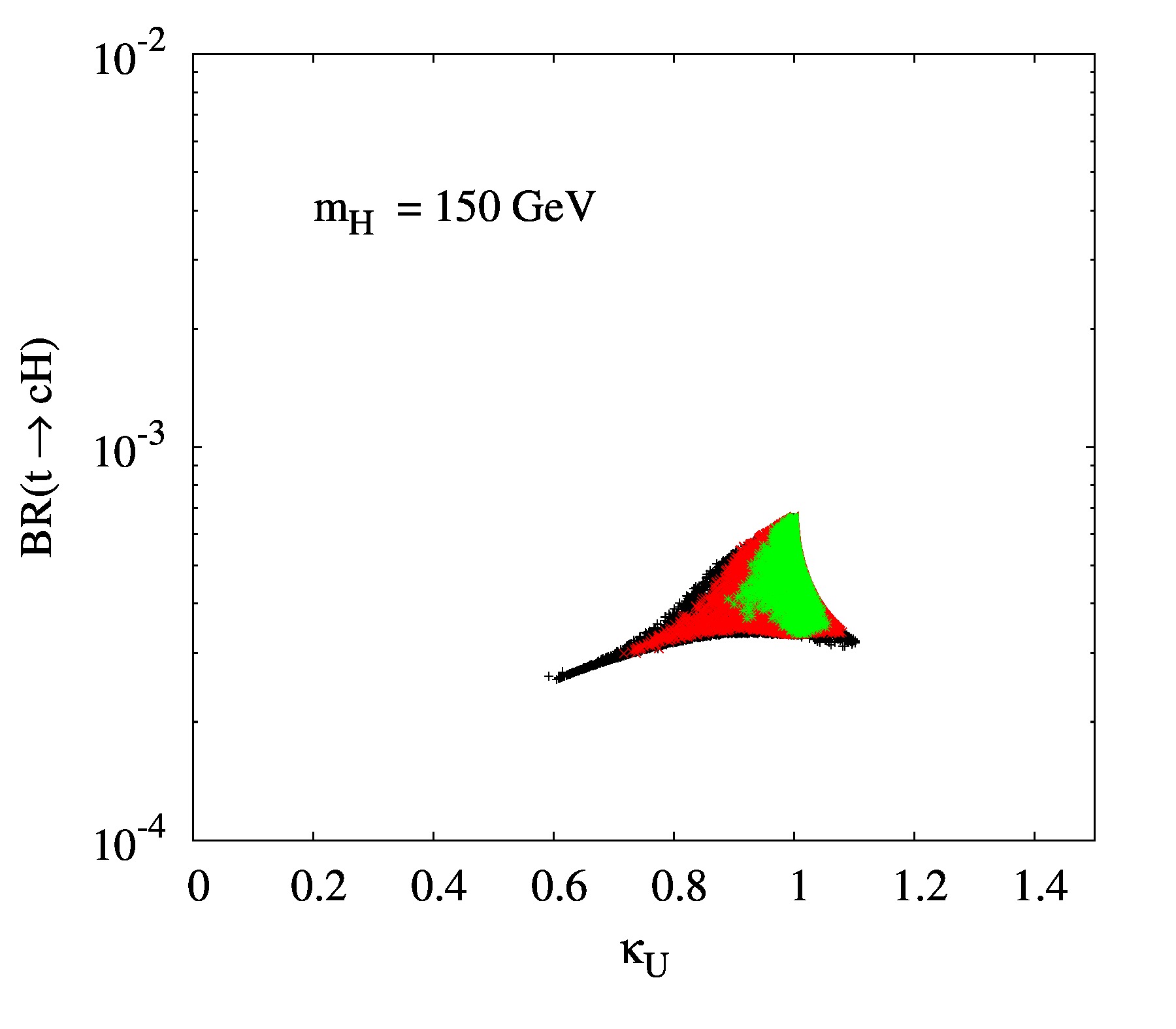}
\includegraphics[width=4.8cm,height=6cm]{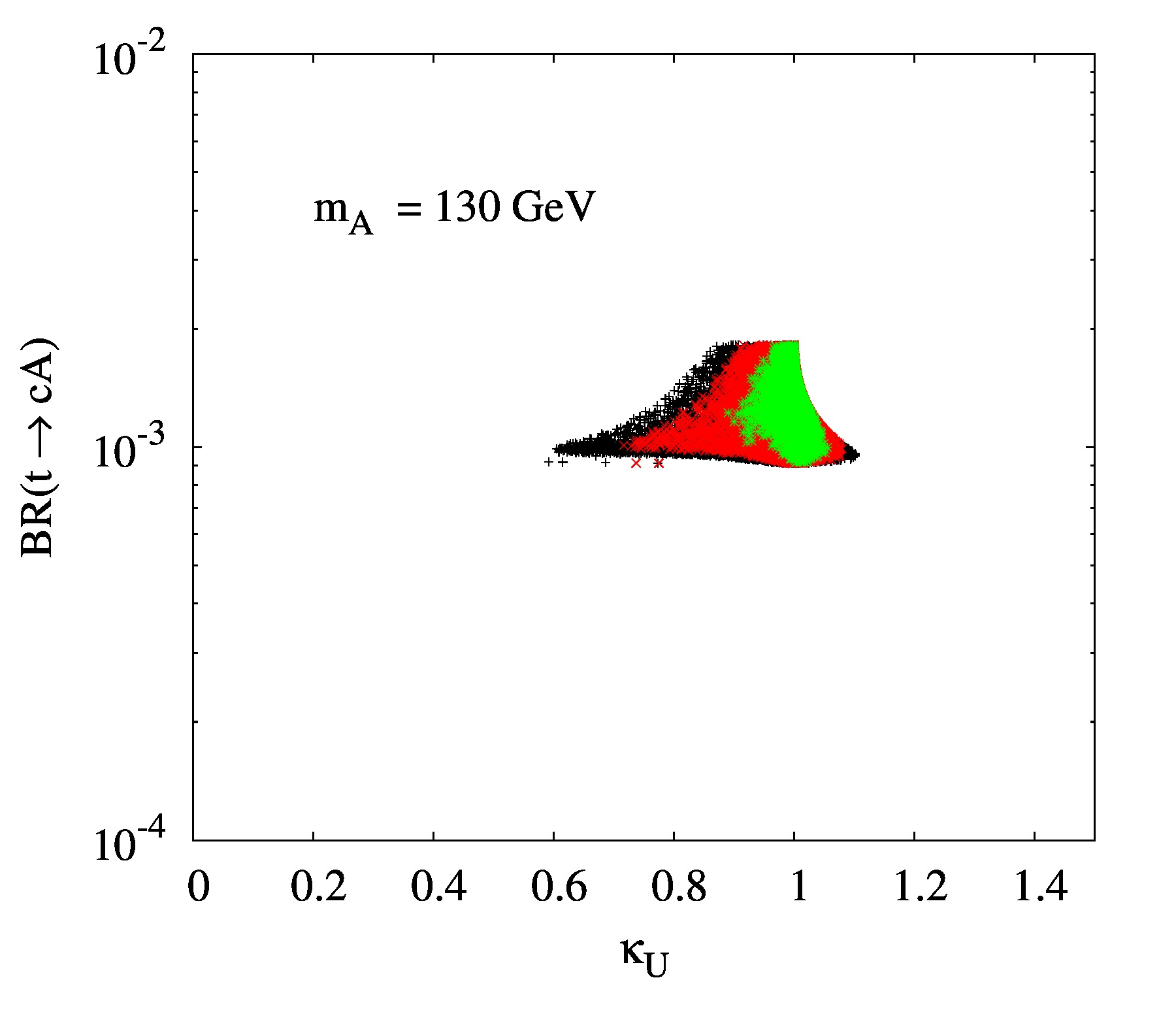}
\caption{Branching ratios of Br$(t\to ch)$(left),  Br$(t\to cH)$(middle)
   and Br$(t\to ch)$(right) as a function of
    $\kappa_U$ in type III with $\chi_{F} = +1$. For $m_h=125.36$ GeV, 
$m_H = 150$ GeV and $m_A = 130$ GeV.}
\label{fig:brtch1}
\end{figure}

\begin{figure}[ht!]
\includegraphics[width=4.8cm,height=6cm]{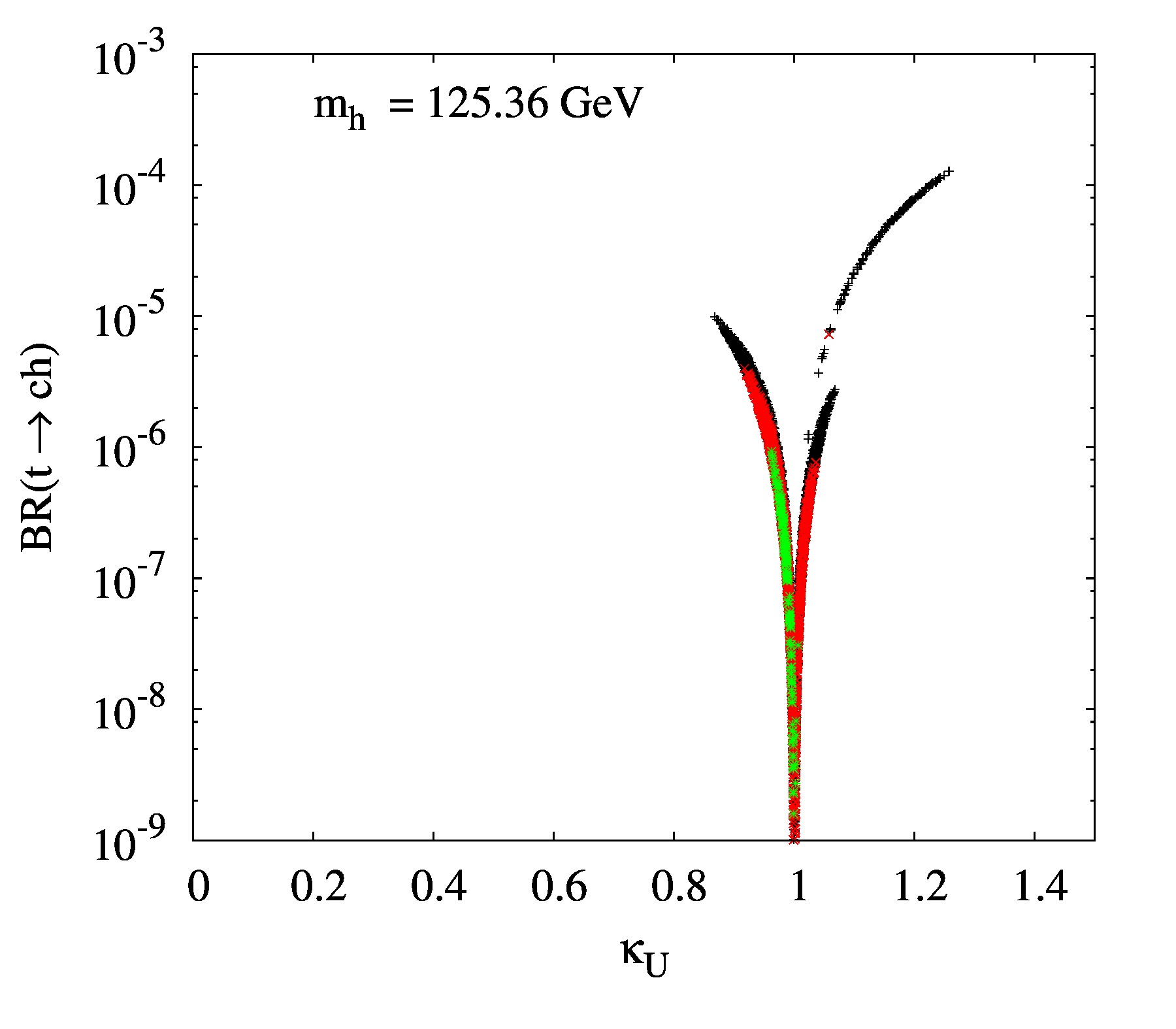}
\includegraphics[width=4.8cm,height=6cm]{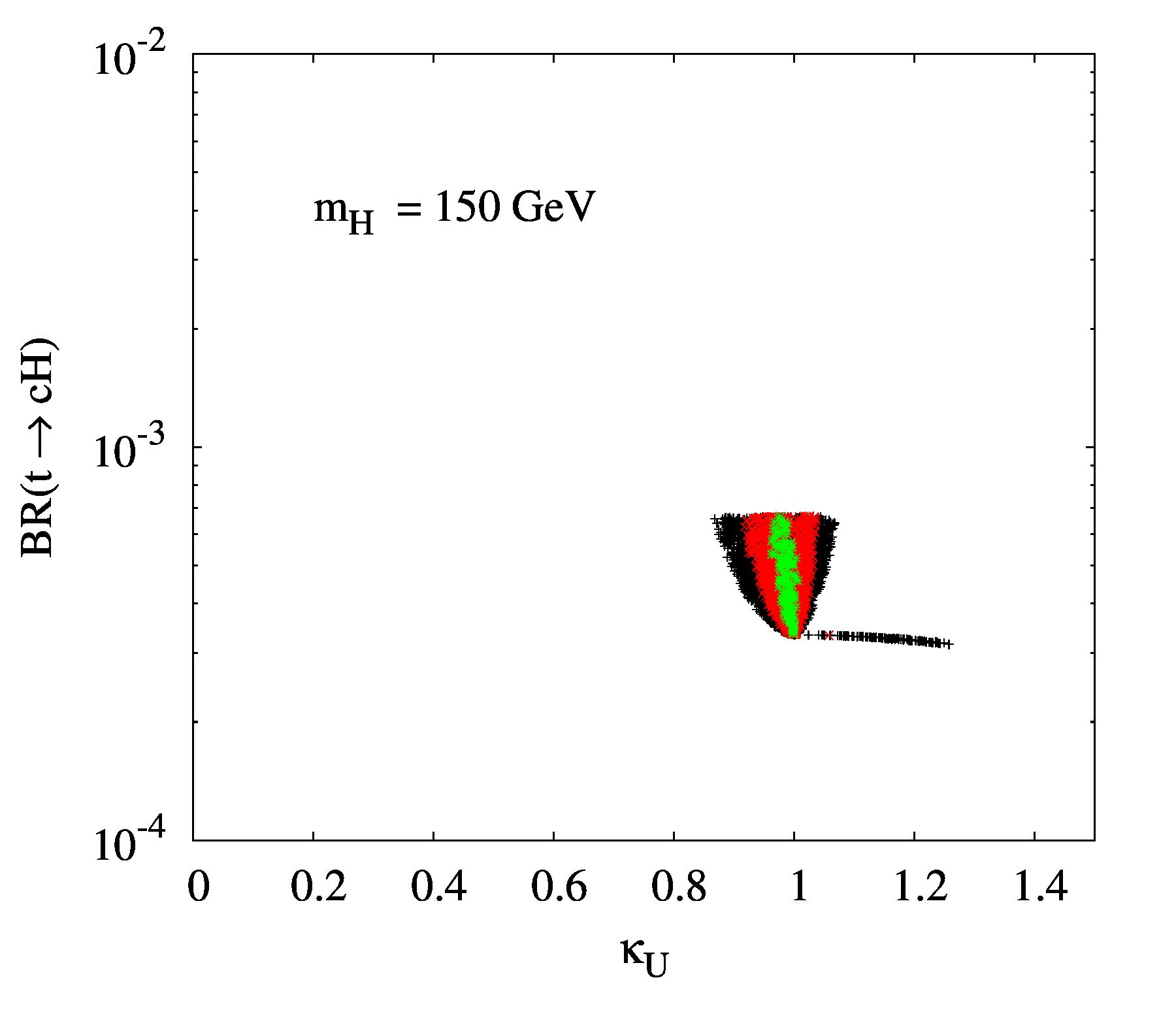}
\includegraphics[width=4.8cm,height=6cm]{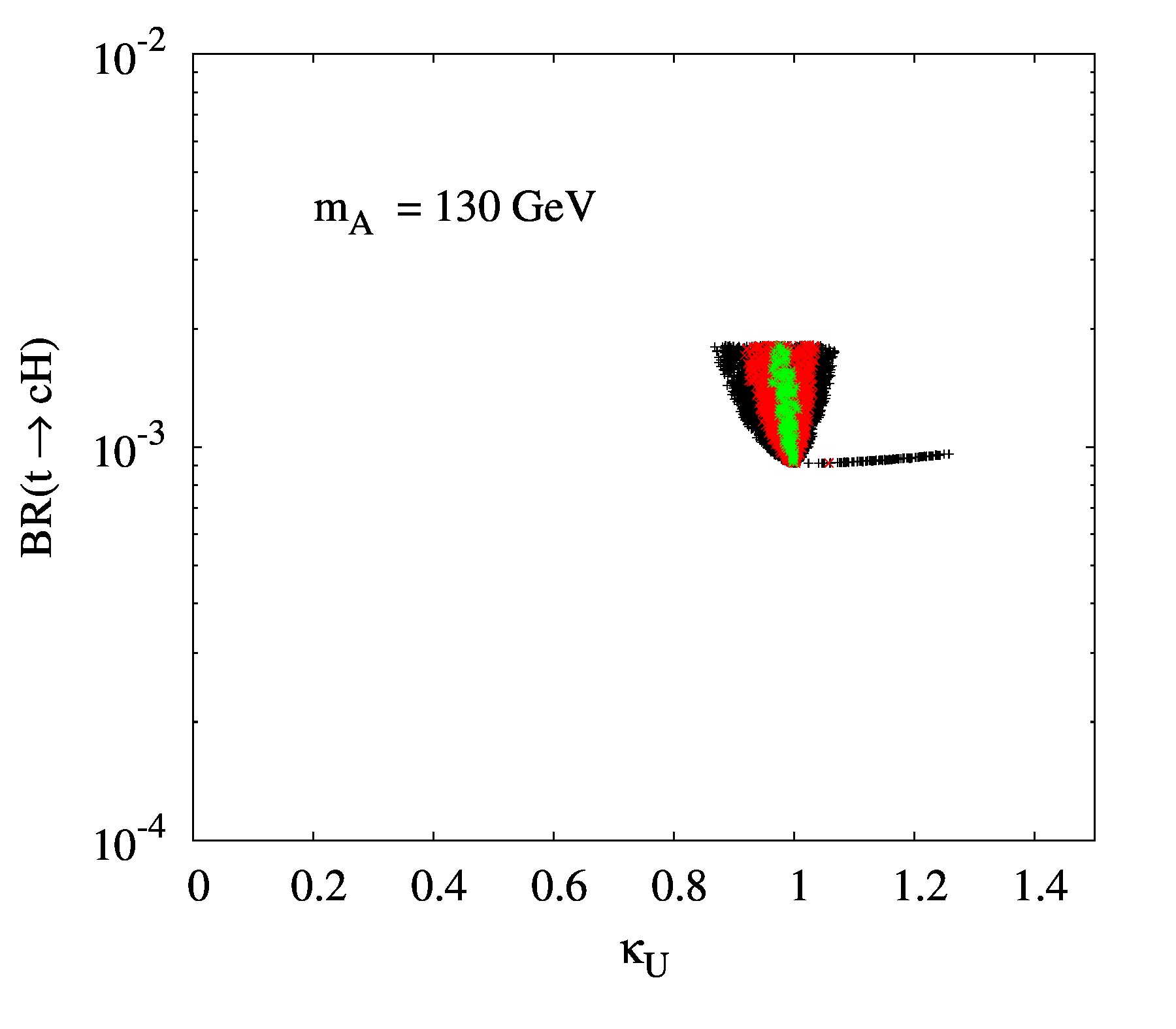}
\caption{The legend is the same as  Fig.~\ref{fig:brtch1} but for $\chi_{F} = -1$. }
\label{fig:brtch2}
\end{figure}

In Fig. \ref{fig:brtch1} and \ref{fig:brtch2}, we show the fitted branching
fractions for $t\to ch$ (left), $t\to cH$ at $m_H=150$ GeV (middle) and  $t\to cA$ at $m_A=130$ GeV (right) as a function of $\kappa_U$, where 
 Fig.~\ref{fig:brtch1} is for $\chi_{F} = +1$ while Fig.~\ref{fig:brtch2} is $\chi_{F}=-1$.  In the case of $\chi_{F} = +1$
 the fitted value for $\kappa_U$ at the 3$\sigma$ level 
is in the range $[0.6,1.18]$ and the branching fraction for 
$t\to ch , cH$ are less than $10^{-3}$ while $Br(t\to cA)$ slightly exceed 
 the $10^{-3}$ level.  Similarly, for $\chi_{F} = -1$
 the fitted value for $\kappa_U$ at the 3$\sigma$ level 
is in the range $[0.85,1.25]$ and the branching fraction for 
$t\to ch , cH, cA$ are the same size as in the previous case.
%
\section{Conclusions}

For studying the constraints of $8$ TeV LHC experimental data, we perform  $\chi$-square analysis to find the  most  favorable 
regions for the free parameters in the two-Higgs-doublet models. For comparisons, we focus on the type-II and type-III models, in which the latter model not only affects the flavor conserving Yukawa couplings, but also generates the scalar-mediated flavor-changing neutral currents at the tree level. 

Although the difference between type-II and type-III is the Yukawa sector, however, since the new Yukawa couplings in type-III are  associated with $\cos(\beta-\alpha)$ and $\sin(\beta-\alpha)$, the modified couplings $tth$ and $btH^{\pm}$ will change the constraint of free parameters. 

In order to present the influence of modified Yukawa couplings, we show the allowed values of $\sin\alpha$ and $\tan\beta$ in Fig.~\ref{fig:satb}, where the LHC updated data for $pp\to h \to f$ with $f=\gamma\gamma$, $WW^*/ZZ^*$ and $\tau^+\tau^-$ are applied and other bounds are also included. By the results, we see that $\sin\alpha$ and $\tan\beta$ in type-III gets even stronger constraint if the dictated parameter $\chi_F = -1$ is adopted; on the contrary, if we take $\chi_F=+1$, the allowed values for $\sin\alpha$ and $\tan\beta$ are wider. 
It has been pointed out that there exist the wrong sign Yukawa couplings to down type quarks in the type-II model, i.e. $\sin\alpha >0$ or $\kappa_D <0$. By the study, we find that except the allowed regions of parameters are shrunk slightly, the situation in $\chi_F=-1$ is similar to the type-II case. In $\chi_F=+1$, although the $\kappa_D <0$ is not excluded completely  yet, but the case has a strict limit by current data. We show the analyses in Figs.~\ref{fig:cdcu} and \ref{fig:cdcv}. In these figures, one can also see the correlations with modified Higgs coupling to top-quark  $\kappa_U$ and to gauge boson $\kappa_V$. 

When the parameters are bounded by the observed channels, we show the
influence on the unobserved channel $h\to Z\gamma$ by using the scaling factor
$\kappa_{Z\gamma}$, which is defined by the ratio of decay rate to the SM
prediction. We find that the change of $\kappa_{Z\gamma}$ in type-III with
$\chi_F=-1$ is less than $10\%$; however, with $\chi_F=+1$, the value of
$\kappa_{Z\gamma}$ could be lower from SM prediction by over $10\%$. We also 
show our predictions for signal strengths $\mu_{\gamma\gamma}$ and 
$\mu_{\gamma Z}$ and their correlation at 13 TeV.

The main difference between type-II and -III model is:  the flavor changing neutral currents in the former are only induced by loops, while in the latter they could occur at the tree level. We study the scalar-mediated $t\to c (h, H, A)$ decays in type-III model and find that when all current experimental constraints are considered, $Br(t\to c(h, H) )< 10^{-3}$ for $m_h=125.36$ and $m_H=150$ GeV and  $Br(t\to cA)$ slightly exceeds $10^{-3}$ for $m_A =130$ GeV. The detailed numerical analyses are shown in Figs.~\ref{fig:brtch}, \ref{fig:brtch1} and \ref{fig:brtch2}. 

\section*{Acknowledgments}
The authors thank Rui Santos for useful discussions. A.A would like to thank NCTS for warm hospitality 
where part of this work has been done. 
The work of CHC is supported by the Ministry of Science and
Technology of R.O.C under Grant \#: MOST-103-2112-006-004-MY3. 
The work of M. Gomez-Bock was partially supported by  UNAM under  PAPIIT IN111115.
This work was also supported by the Moroccan Ministry of Higher
Education and Scientific Research MESRSFC and  CNRST: "Projet dans les
domaines prioritaires de la recherche scientifique et du 
d\'eveloppement technologique": PPR/2015/6.



\end{document}